%% file: main.tex
\documentclass[aip, jcp, reprint, twocolumn, tightenlines, superscriptaddress, floatfix]{revtex4-1}

\input{preamble}

\begin{document}



\title{Uncertainty in the era of machine learning for atomistic modeling}

\author{Federico Grasselli}
\affiliation{Dipartimento di Scienze Fisiche, Informatiche e Matematiche, Universit\`a degli Studi di Modena e Reggio Emilia, 41125 Modena, Italy}
\affiliation{CNR NANO S3, 41125 Modena, Italy}
\email{federico.grasselli@unimore.it}

\author{Sanggyu Chong}
\affiliation{Laboratory of Computational Science and Modeling, Institute of Materials, \'Ecole Polytechnique F\'ed\'erale de Lausanne, 1015 Lausanne, Switzerland}

\author{Venkat Kapil}
\affiliation{Yusuf Hamied Department of Chemistry, University of Cambridge, Cambridge CB2 1EW, United Kingdom}
\affiliation{Department of Physics and Astronomy, University College London, London, United Kingdom}
\affiliation{Thomas Young Centre and London Centre for Nanotechnology, University College London, London WC1E 6BT, United Kingdom}

\author{Silvia Bonfanti}
\affiliation{Center for Complexity and Biosystems, Department of Physics ``Aldo Pontremoli'', University of Milan, Via Celoria 16, 20133 Milano, Italy}
\affiliation{NOMATEN Centre of Excellence, National Center for Nuclear Research, ul. A. So\l{}tana 7, 05-400 Swierk/Otwock, Poland}

\author{Kevin Rossi}
\affiliation{Department of Materials Science and Engineering, Delft University of Technology, 2628 CD, Delft, The Netherlands}
\affiliation{Climate Safety and Security Centre, TU Delft The Hague Campus, Delft University of Technology, 2594 AC, The Hague, The Netherlands}
\email{k.r.rossi@tudelft.nl}

\begin{abstract}
The widespread adoption of machine learning surrogate models has significantly improved the scale and complexity of systems and processes that can be explored accurately and efficiently using atomistic modeling. However, the inherently data-driven nature of machine learning models introduces uncertainties that must be quantified, understood, and effectively managed to ensure reliable predictions and conclusions.
Building upon these premises, in this Perspective, we first overview state-of-the-art uncertainty estimation methods, from Bayesian frameworks to ensembling techniques, and discuss their application in atomistic modeling. 
%
%
We then examine the interplay between model accuracy, uncertainty, training dataset composition, data acquisition strategies, model  transferability, and robustness. 
In doing so, we synthesize insights from the existing literature and highlight areas of ongoing debate. 

%
%

\end{abstract}

\maketitle

\section{Introduction}


Tycho Brahe, 16th century Danish astronomer, is credited for the ``great care he took in correcting his observations for instrumental errors'',\cite{vinter1942tycho} introducing the concept of measurement-theory inconsistency in astronomy, thus turning it into an empirical science.
Since then, the ability to assess instrument and model errors as well as quantify the uncertainty and confidence intervals when making predictions
has become a pillar of the scientific method and, in fact, discriminates between what is scientific and
what is not. 


In many cases, chemists and materials scientists draw conclusions based on incomplete or uncertain information, as it is often the case when dealing with expensive, time-consuming, oftentimes noisy measurements. Uncertainty quantification (UQ) provides a framework for systematically incorporating uncertainty in this scientific process, thereby enhancing the reliability, robustness, and applicability of experimental and theoretical results. 
In the context of materials science, chemistry, and condensed matter physics,  researchers optimize materials and properties while accounting for uncertainties, variations, and errors in their measurements and theories (e.g., via replication and sensitivity analysis). This improves the reliability and validity in the models of physical phenomena and design of novel materials and processes.

Nowadays, machine learning and artificial intelligence methods are emerging as a key tools for accelerating the design, engineering, characterization, and understanding of materials, molecules, and reactions at interfaces
In the context of atomistic modeling, machine learning facilitates the development of predictive models and interatomic potentials that can simulate material behavior with high accuracy and reduced computational cost compared to traditional methods. Incorporating UQ into these machine learning models is crucial for assessing the reliability of predictions and understanding the limitations of the models.
By quantifying uncertainties, researchers can identify areas where the model's predictions are less certain, guide the selection of new data points for training (active learning), understand how to train robust models, and make informed decisions about the deployment of these models in practical applications. 

\begin{figure}
    \centering
    \includegraphics[width=1\linewidth]{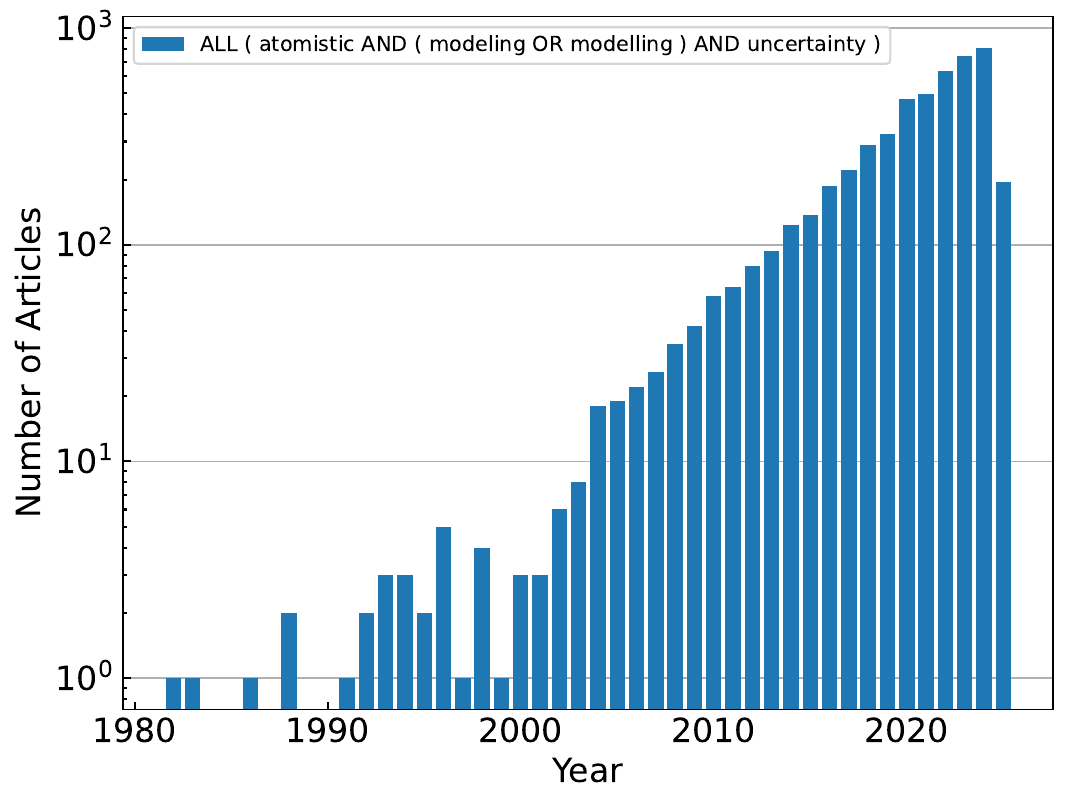}
    \caption{Frequency (count) of articles per year corresponding to the search query reported in the figure legend: all fields of research, querying atomistic \& model(l)ing \& uncertainty. All fields of research refers to articles titles, abstracts, keywords, and references. Data retrieved from Scopus accessed on 06 Mar 2025. }
    \label{fig:count}
\end{figure}

In this perspective, we examine how the integration of machine learning and UQ enhances the predictive capabilities of atomistic simulations and ensures that inherent uncertainties are systematically accounted for, leading to more robust and reliable materials design and discovery. 
%
In this context, we acknowledge the recent contributions to the topic by \textcite{Dai2024} and \textcite{kulichenko2024}.
By the same token, we remark that, while we aim for a complete discussion, we intentionally focus on a a relatively restricted number of representative works in the literature.
A search for the term ``Atomistic Modeling (or Modelling)'' and ``Uncertainty'' shows that these appear with an increasing frequency, amounting to more than 5000 literature items (Figure \ref{fig:count}), highlighting the significance of the topic under scrutiny.

The focus of our work is then on analysing recent trends in uncertainty estimation methods —such as Bayesian frameworks and ensemble approaches — and their practical application to assess prediction reliability, and guide efficient data acquisition.
When possible, 
we synthesize and unify insights from the literature.
Examples include connection interpretation of uncertainty and extrapolation measures as Mahalanobis distances and discussion on geometrical and statistical approaches to define in- and out-of distribution predictions.
When consensus is lagging, we highlight key gaps and open research questions.
These concern benchmarking of uncertainty quantification methods and models trasnferability, uncertainty propagation for dynamical observables; uncertainty quantification in data-efficient methods including foundational and multi-fidelity approaches.
In conclusion, we aim to provide a unified framework and highlight the open questions toward robust, efficient, and interpretable machine learning approaches for atomistic modeling.

\section{Uncertainty Estimates}\label{sec:uncertainty_estimate}

For an uncertainty quantification method to be effective, a number of properties are desirable. In particular, the UQ methods shall be:\cite{Dai2024}
\begin{enumerate}
    \item \emph{accurate}, by realistically modeling the true uncertainty associated with the ML prediction, and aiming to minimize bias and systematic errors;
    \item \emph{precise} enough to provide a sufficiently narrow range of possible values; 
    \item \emph{robust}, against variations in the data or model assumptions, providing reliable results also when tested out-of-domain;
    \item \emph{traceable and comprehensive}, by capturing and identifying all the possible sources of uncertainty, which include the choice of hyperparameters and training set data points, or the stochastic optimization of non-deterministic models.
    \item \emph{computationally efficient}, requiring only a negligible overhead, possibly also in training, in obtaining the uncertainty values of interest from the ML model
\end{enumerate}

In what follows, we adopt operative definitions of uncertainty based on the variance (second moment) of the distribution of predictions (either theoretical or constructed via ensembles) to quantify the spread of uncertain outcomes. 
This definition indeed displays the properties listed above.
The analysis of first and second moment only may not be fully descriptive for non-Gaussian (e.g., skew, heavy-tailed or multi-modal) distributions. 
Nonetheless, it provides an interpretable and computationally lightweight measure of variability. 
Furthermore, it aligns well with Gaussian or near-Gaussian models, such as those that are built from the Laplace approximation (see Sec. \ref{sec:NN}).
Finally, it supports simple calibration strategies that leverage the comparison of the uncertainty estimate with the second moment of the empirical distribution followed by the residuals $y_i - \tilde{y}_i$ between the reference value for input $i$ and its ML prediction.

%
%
Finally, towards a clear and unified discussion, we spell out the notation we will adopt for our successive considerations: 

\begin{itemize}
    \item[$\mathbf{x}$] (or $\mathbf{x}_i$ when labeling is needed) generic input/sample
    \item[$\mathbf{X}$] matrix collecting inputs in the training set as rows
    \item[$\mathbf{f}_i$] array of features corresponding to $\mathbf{x}_i$
    \item[$\mathbf{F}$] matrix collecting training-set features as rows
    \item[$\mathbf{w}$] parameters of the model (a.k.a. weights)
    \item[$\tilde{y}_i$] $\equiv \tilde{y}(\mathbf{x}_i)$ machine-learning prediction for input $\mathbf{x}_i$
    \item[$y_i$] reference value corresponding to input $\mathbf{x}_i$
    \item[$\mathcal{D}$] training dataset of input-label pairs $(\mathbf{x}_i,y_i)$, with $i=1,\ldots,N_\mathrm{train}$
    \item[$\sigma^2_i$] variance on prediction $\tilde{y}_i$
    \item[$\alpha$] calibration constant (see Sec.~\ref{sec:calib})
    \item[$\mathcal{L}$] training-set loss function
    \item[$\ell_i$] term of the loss function corresponding to a single instance $i$ of the training set
    \item[$\tilde{\sigma}(\mathbf{x})$] machine-learning prediction, corresponding to input $\mathbf{x}$, for the uncertainty in mean-variance estimates and mean-variance ensembles, see Sec. \ref{subsec: MVE est and ens}. 
\end{itemize}


%

\subsection{Formulae for  direct (and simple) uncertainty estimates}\label{sec:direct_estimates}

\begin{figure}
    \centering
    \includegraphics[width=7cm]{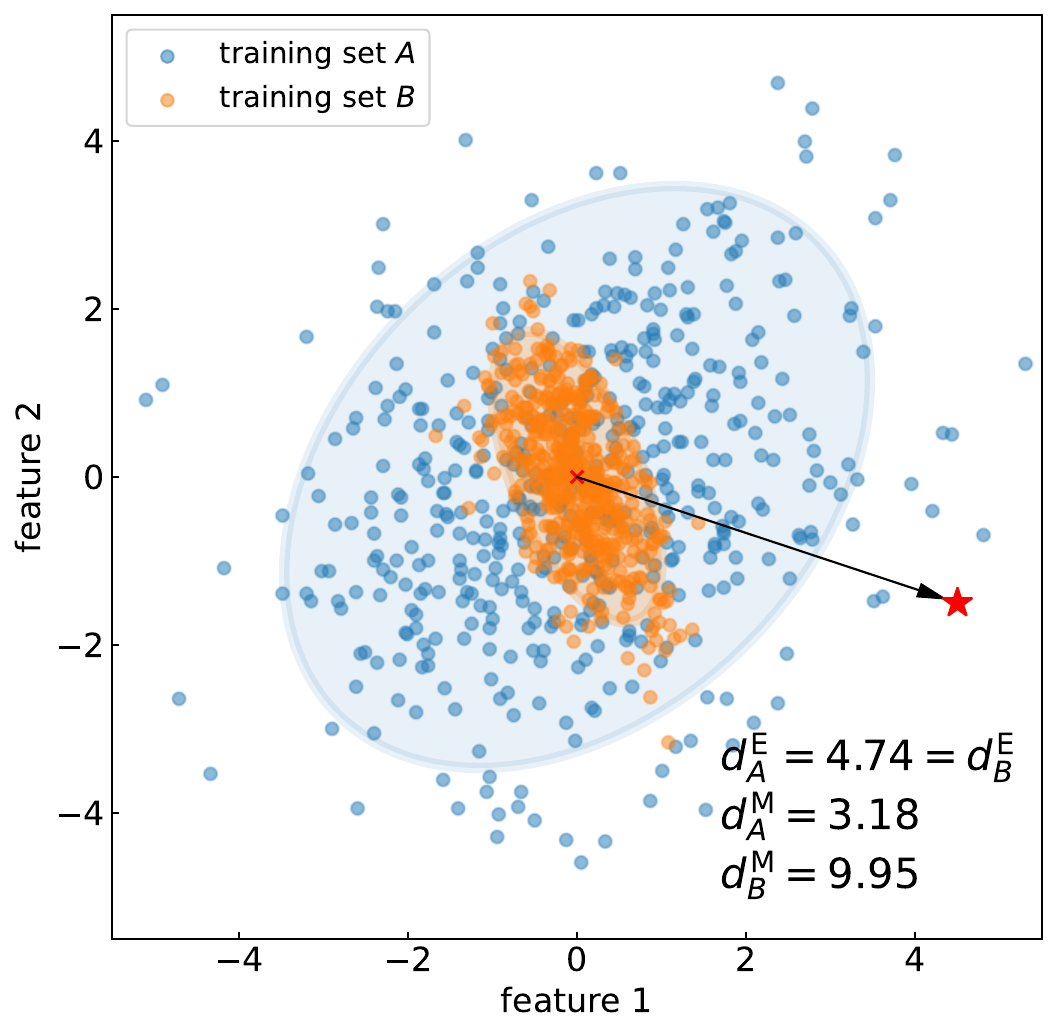}
    \caption{Mahalanobis distance. The new sample $\star$ has equal Euclidean distance $d^\mathrm{E}$ between the distribution of features in training set $A$ (blue), characterized by a large covariance, and the distribution of features in training set $B$ (orange), characterized by a smaller covariance. The shaded ellipses have axes equal to the eigenvalues of the covariance. In striking contrast to Euclidean distance, the Mahalanobis distance of $\star$ from the distribution of features in training set $B$, $d^\mathrm{M}_B$, is more than three times larger than that from $A$, $d^\mathrm{M}_A$.}
    \label{fig:Mahalanobis}
\end{figure}

In the literature there exist several direct formulae to estimate the ML uncertainty on a given prediction, which make the details much dependent on the specific ML approach, e.g., linear/kernel ridge regression; full or sparse Gaussian process regression (GPR); neural-network (NN) models. Nonetheless, all these direct estimates share a common (Bayesian) interpretation. In fact, for a given new sample $\star$, the general shape of the uncertainty associated to the prediction of is in the form of a Mahalanobis (square) distance:\cite{Mahalanobis1936}
\begin{equation}
    \sigma_\star^2 = \alpha^2 \mathbf{f}_\star^\top \mathbf{G}
    \mathbf{f}_\star \label{eq:sigma2star}
\end{equation}
i.e.~the (non-Euclidean) norm of properly defined feature vector, $\mathbf{f}_\star$,  that the model associates to the new sample (see also Figure \ref{fig:Mahalanobis} for additional insights). 
For simplicity, we assume the features have been centered, i.e.~that $\frac{1}{N_\mathrm{train}}\sum_{i=1}^{N_\mathrm{train}} f_{i,a} = 0, \forall a = 1, \ldots, N_f$. The prefactor $\alpha^2$ is independent of $\star$ and acts as a tuneable constant that must calibrated on some validation dataset (and also provides the correct units for the variance of the predictions). Why calibration is needed and how to calibrate uncertainty are discussed in Sec.~\ref{sec:calib}. The shape of the positive-definite metric tensor $\mathbf{G}$ is model-dependent, but possess some common characteristics:
\begin{enumerate}
    \item it can be viewed as an inverse covariance matrix of the properly defined features of the input data points in $\mathcal{D}$, i.e.~$\mathbf{G} = [\mathrm{cov}(\mathbf{F})]^{-1}$, where $\mathbf{F}\in \mathbb{R}^{N_\mathrm{train} \times N_f}$ collects as rows the transpose of the feature vectors $\{\mathbf{f}_i\}_{i=1,\ldots,N_\mathrm{train}}$ of the training points;
    \item it is therefore strongly dependent on the distribution of input points $\mathbf{x}_i$ in $\mathcal{D}$ and on how much the new point $\star$ is ``close'' to such distribution in this metric space;
    \item it is largely independent of the specific target values $y_i$ in $\mathcal{D}$.\footnote{The only dependence on the specific target quantity and values is through the value of the regularizer that is included to make the inversion of the covariance matrix numerically stable.}
\end{enumerate}

We report below the specific, model-dependent expression of the features $\mathbf{f}$ and therefore of the metric tensor $\mathbf{G}$. 

\subsubsection{Linear regression}
In a linear regression $\mathbf{f} \equiv \mathbf{x}$, so that $N_f = D$, and
\begin{equation}
    \tilde{y}(\mathbf{x}, \mathbf{w}) = \mathbf{x}^\top \mathbf{w}
\end{equation}
where $\mathbf{w}$ are the weights. In a Bayesian picture, if  we assume the weights to be sampled from a zero-mean Gaussian prior, we have:
\begin{equation}
    \mathbf{G} = \left(\mathbf{X}^\top \mathbf{X} + \varsigma^2 \mathbf{I}_{D}\right)^{-1} 
\end{equation}
where $\varsigma^2$ acts as a regularizer strength and $\mathbf{I}_D$ is the identity matrix of size $D$ equal to the number of components of (any) input $\mathbf{x}\in \mathbb{R}^D$. Therefore, the uncertainty on a prediction $\tilde{y}_\star = \tilde{y}(\mathbf{x}_\star)$ is
\begin{equation}
    \sigma_\star^2 = \alpha^2 \mathbf{x}_\star^\top \left(\mathbf{X}^\top \mathbf{X} + \varsigma^2 \mathbf{I}_{D}\right)^{-1} \mathbf{x}_\star
\end{equation}

\subsubsection{Gaussian process regression}
In the GPR problem $\mathbf{f} \equiv \bm{\phi}(\mathbf{x})$, i.e.~the regression exercise has the form
\begin{equation}
    \tilde{y}(\mathbf{x}) = [\bm{\phi}(\mathbf{x})]^\top \mathbf{w},\label{eq:yGPR}
\end{equation}
where $\bm{\phi}(\mathbf{x})$ maps the $D$-dimensional input $\mathbf{x}$ into a $N_f$-dimensional feature space (in general, as a nonlinear function of the input). The components $\phi^a(\mathbf{x})$, with $a=1,\ldots,N_f$ are often called \textit{basis functions}.

By assuming again that the weights are sampled from a zero-mean Gaussian prior, we have:
\begin{equation}
    \mathbf{G} = \left(\bm{\Phi}^\top \bm{\Phi} + \varsigma \mathbf{I}_{N_f}\right)^{-1} 
\end{equation}
and the uncertainty on a prediction $\tilde{y}_\star = \tilde{y}(\mathbf{x}_\star)$ is
\begin{equation}
    \sigma_\star^2 = \alpha^2 \bm{\phi}(\mathbf{x}_\star)^\top \left(\bm{\Phi}^\top \bm{\Phi} + \varsigma \mathbf{I}_{N_f}\right)^{-1} \bm{\phi}(\mathbf{x}_\star)
    \label{eq:GPR_cov}
\end{equation}

Oftentimes the reported GPR uncertainty formula is:
\begin{equation}
\begin{split}
    &\sigma_\star^2 = k(\mathbf{x}_\star, \mathbf{x}_\star) \\
    &- \mathbf{k}(\mathbf{x}_\star, \mathbf{X})^\top
    \left[\mathbf{K}(\mathbf{X}, \mathbf{X}) + \varsigma \mathbf{I}_{N_\mathrm{train}}
    \right]^{-1}
    \mathbf{k}(\mathbf{x}_\star, \mathbf{X})
    \label{eq:GPR_ker}
\end{split}
\end{equation}
which makes use of the kernel that, for any pair of inputs $\mathbf{x}_i$ and $\mathbf{x}_j$, is  defined as $k(\mathbf{x}_i, \mathbf{x}_j) \propto [\bm{\phi}(\mathbf{x}_i)]^\top \bm{\phi}(\mathbf{x}_j)$. 
To switch from Eq.~\eqref{eq:GPR_cov} to Eq.~\eqref{eq:GPR_ker} or vice versa, it is sufficient to use Woodbury's identity, after assuming all the needed matrix inversions are possible. In this sense, the role of the regularizer is crucial: in fact, whenever $\bm{\Phi}$ is not full rank, either $\bm{\Phi}^\top \bm{\Phi}$ is invertible and $\bm{\Phi} \bm{\Phi}^\top$ is not (case of ``tall'' matrix $\bm{\Phi}$, with $N_\mathrm{train} > N_f$), or $\bm{\Phi} \bm{\Phi}^\top$ is invertible and $\bm{\Phi}^\top \bm{\Phi}$ is not (case of ``broad'' matrix $\bm{\Phi}$, with $N_\mathrm{train} < N_f$). For the relation between the eigenvalues/-vectors of $\bm{\Phi}^\top \bm{\Phi}$ and  $\bm{\Phi} \bm{\Phi}^\top$, see \textcite{tipping2000sparse}.

Mercer's theorem ensures that for any kernel there exist a possibly infinite (i.e.~$N_f\to \infty$) set of basis functions. Furthermore, the representation in terms of the functions $\bm{\phi}$ is also very useful when sparse kernel approximations, such as the Nystr\"om method detailed in Appendix \ref{appendix:Nystrom}, are employed.

\subsubsection{Neural networks}\label{sec:NN}

Expressions analogous to Eq.~\eqref{eq:sigma2star} have appeared for neural networks several decades ago, in the work of  MacKay in the early '90s \cite{mackay1992bayesian,mackay1992practical,mackay1992information}, introducing Laplace approximation within the context of Bayesian approach to neural networks. 
The Laplace approximation consists in approximating the posterior distribution of the weights as a multivariate Gaussian distribution, centered around the maximum a posteriori (MAP) \textit{optimal} weights $\mathbf{w}_o$, that are obtained after the NN training:
\begin{equation}
        p(\mathbf{w}|\mathcal{D}) \approx \mathcal{N} \big(\mathbf{w}_o, \alpha^2 \mathbf{G}\big) \label{eq:Laplace_approx}
\end{equation}
where $\alpha^2 \mathbf{G}$ is the covariance matrix of the weights close to MAP, and the tuneable parameter $\alpha$ may be interpreted as a noise level on observation \cite{mackay1992information}. The discrepancy principle would indicate the mean square error of the observations \cite{cohn1996neural} as an empirical estimate of $\alpha^2$; nonetheless, the latter is often treated as a tuneable parameter, since  additional calibrations are often required, as reported in, e.g.,~\textcite{mackay1992bayesian} and \textcite{imbalzano2021uncertainty}.   
In the Laplace approximation, the matrix $\mathbf{G}$ is given by the inverse Hessian matrix of  loss function $\mathcal{L} = \sum_{i} \ell_i(\tilde{y}(\mathbf{x}_i, \mathbf{w}), y_i)$, computed at MAP:
\begin{equation}
    \mathbf{H}_o = \left.\frac{\partial^2 \mathcal{L}}{\partial \mathbf{w} \, \partial \mathbf{w}^\top} \right|_{\mathbf{w}_o} = \sum_{i=1}^{N_\mathrm{train}} \left.\frac{\partial^2 \ell_i}{\partial \mathbf{w} \, \partial \mathbf{w}^\top} \right|_{\mathbf{w}_o}
    \label{eq:Hessian_MAP}
\end{equation}
$\mathbf{H}_o$ is routinely approximated by its Gauss-Newton form, which employs only first order derivatives of predictions with respect to the weights and evaluated at MAP, $\bm{\phi}_i \equiv \left.\frac{\partial \tilde{y}_i}{\partial \mathbf{w}}\right|_{\mathbf{w}_o}^\top$ which can be easily retrieved by backpropagation:
\begin{equation}
    \mathbf{H}_o \approx \sum_{i=1}^{N_\mathrm{train}} \bm{\phi}_i \, \frac{\partial^2 \ell_i}{\partial \tilde{y}_i^2} \, \bm{\phi}_i^\top
    \label{eq:Gauss-Newton_Hessian}
\end{equation}
Finally, the distribution of the output corresponding to the input $\star$ becomes:\cite{daxberger2021laplace}
\begin{equation}
    p(y_\star | \mathbf{x}_\star, \mathcal{D}) \approx \mathcal{N}\left(
    \Tilde{y}(\mathbf{x}_\star,\mathbf{w}_o); \alpha^2 \, \bm{\phi}_\star^\top \mathbf{H}_o^{-1}  \bm{\phi}_\star
    \right)\label{eq:distr_output_NN_LA}
\end{equation}
from which the variance $\sigma_\star^2 = \alpha^2 \, \bm{\phi}_\star^\top \mathbf{H}_o^{-1}  \bm{\phi}_\star$ is obtained in the form of a Mahalanobis distance, Eq.~\eqref{eq:sigma2star}. The same results can be obtained in an alternative but equivalent mathematical construction probing how robust a ML model is to a change in the prediction of an input $\star$, based on a constrained minimization of the loss.\cite{chong2023robustness,bigi2024prediction} (see Sec.~\ref{sec:pred_rigidity} for more details).

Two remarks should be made: first,
as it is reasonable to expect, the quality of this approximation depends on how much the posterior distribution of the NN model is close to a multivariate Gaussian; second, the large number of weights, and thus of components in $\bm{\phi}_i$, in current deep NN typical architectures makes the storage of $\mathbf{H}_o$ unfeasible, even in the Gauss-Newton approximation, due to memory requirements quadratic in the size of the $\bm\phi$ arrays. The context of NN Gaussian processes,\cite{lee2017deep}and in particular of the Neural Tangent Kernel formalism,\cite{jacot2018neural,lee2019wide} provides the ideal theoretical framework to justify the first point and to find a viable strategy to overcome the second one. In  ~\textcite{bigi2024prediction}, the use of a last-layer (LL) approximation was extensively justified, whereby only the derivatives of the predictions with respect to the LL weights $\mathbf{w}^L$, i.e.~the LL latent features
\begin{equation}
    \mathbf{f}_i \equiv \left.\frac{\partial \tilde{y}_i}{\partial \mathbf{w}^L}\right|_{\mathbf{w}_o}^\top
\end{equation}
are considered in building $\mathbf{H}_o$ and in evaluating the variance of the prediction for a new input $\star$. For a mean square loss function, the latter becomes:\footnote{To avoid notation overburden, we used the same symbol $\alpha^2$ in both Eqs.~\eqref{eq:distr_output_NN_LA} and \eqref{eq:LLsigma2star} for the calibration parameters, although there is no reason for them to be equal.}
\begin{equation}
    \sigma_\star^2 = \alpha^2 \mathbf{f}_\star^\top \left(\mathbf{F}^\top \mathbf{F} + \varsigma^2\mathbf{I}_{N_L} \right)^{-1} \mathbf{f}_\star\label{eq:LLsigma2star}
\end{equation}
where a regularizer has been added for numerical stability and where $N_L$ is the number of components of LL latent features, i.e.~the number of nodes of the last hidden layer of the NN. 
A few further remarks: 
\begin{enumerate}
    \item The presented derivation is based on the assumption that no additional nonlinear activation is applied to the product of the LL latent features and LL weights, i.e. that $\tilde{y} = \mathbf{f}^\top \mathbf{w}^L$. Things get more complicated whenever a nonlinear application function $\varphi$ is instead applied, $\tilde{y} = \varphi(\mathbf{f}^\top \mathbf{w}^L)$, even though the correct distribution of the prediction can in principle still be sampled (e.g., by Monte Carlo integration).
    \item The LL latent features $\mathbf{f}_i$ do not explicitly depend on the LL weights $\mathbf{w}^{L}$, which are the weights reported to change more during training.\cite{jacot2018neural} As such, the LL latent features, as well as the covariance matrix $\mathbf{F}^\top \mathbf{F}$, are expected to be rather constant during training.\cite{lee2019wide} Furthermore, the different elements of the array of LL latent features are identically distributed at initialization, and centered around zero, for \textit{any} given sample, because the weights (and in particular $\mathbf{w}^{L-1}$) are taken as independent, identically distributed and centered around zero. This implies that the additional enforcement of feature centering should not change the result of the uncertainty estimate. 
    \item Numerical experiments indicate that, while the calibration of $\alpha^2$ is crucial, the regularizer scarcely affects $\sigma_\star^2$ even when the regularizer strength $\varsigma^2$ is varied over several orders of magnitude.
\end{enumerate}

\subsubsection{Bayesian methods beyond the Laplace approximation}

As already discussed, Bayesian UQ scope is to find the probability distribution of the output (a.k.a.~posterior predictive distribution) by means of Bayes' rule
\begin{equation}
p(y_\star|\mathbf{x}_\star,\mathcal{D}) = \int \mathrm{d}\mathbf{w} p(y_\star|\mathbf{x}_\star,\mathbf{w})p(\mathbf{w}|\mathcal{D})
\end{equation}
from which one can quantify the uncertainty as the second moment of the distribution. 
When Laplace approximation is invoked, one can obtain simple expressions like Eq.~\eqref{eq:distr_output_NN_LA}. 
Whenever this is not possible, an explicit sampling of the posterior 
\begin{equation}
\begin{split}
    p(\mathbf{w}|\mathcal{D}) &\propto \exp\left[-\frac{\mathcal{U}(\mathbf{w})}{\mathcal{T}}\right] \\
    \mathcal{U}(\mathbf{w}) &= -\ln \underbrace{\prod_{i=1}^{N_\mathrm{train}}p(y_i|\mathbf{x}_i, \mathbf{w})}_{\mathrm{likelihood},\,p(\mathcal{D}|\mathbf{w})} - \ln \underbrace{p(\mathbf{w})}_\mathrm{prior}
\end{split}
\end{equation}
is needed. Here, $\mathcal{T}$ represents the posterior ``temperature'', introduced as an additional hyperparameter. The true Bayesian posterior is obtained when $\mathcal{T}=1$;\cite{mackay1992bayesian} when $\mathcal{T}<1$ such a ``cold'' posterior is a sharper distribution.\cite{Izmailov2021}
The sampling is usually performed via Markov chain Monte Carlo (MCMC) methods. In order to generate proposals for the Monte Carlo acceptance step, state-of-the-art techniques often leverage Hamiltonian-like dynamics, whereby the parameters $\mathbf{w}$ are evolved according to the ``forces'' $-\bm\nabla_\mathbf{w} \mathcal{U}(\mathbf{w})$.\cite{Neal2012} 
While standard Hamiltonian MCMC methods may be computationally intractable for current NN architectures featuring a huge number of parameters, stochastic-gradient MCMC algorithms have been recently devised and applied to NN ML interatomic potentials,\cite{Thaler2023} which involve data mini-batching and give results comparable to (deep) ensemble methods (see Sec.~\ref{subsec:ensembles_models}).

\subsection{Ensembles of models}\label{subsec:ensembles_models}

Another class of approaches to quantify the ML uncertainty exists, which is based on the generation of an ensemble of  several equivalent models $y^{(m)}(\mathbf{x})$, to compute the empirical mean
\begin{equation}
    \overline{y}(\mathbf{x}) = \frac{1}{M}  \sum_{m=1}^{M} y^{(m)}(\mathbf{x}) \label{eq:comm_mean}
\end{equation}
and variance
\begin{equation}
    \sigma^2(\mathbf{x}) = \frac{1}{M-1} \sum_{m=1}^{M} [y^{(m)}(\mathbf{x}) - \overline{y}(\mathbf{x})]^2
    \label{eq:comm_var}
\end{equation}
for the prediction corresponding to any given input of features $\mathbf{x}$. 


The ensemble members $y^{(m)}(\mathbf{x})$ are routinely obtained in different ways:
\begin{enumerate}
    \item by subsampling the entire dataset and then training one model for each of the subsampled datasets $\mathcal{D}^{(m)}$. The size of $\mathcal{D}^{(m)}$ depends on the subsampling technique, but usually amounts to $\frac{M-1}{M} \times N_\mathrm{train}$.
    
    \item for models that are not trained via a deterministic approach, stochasticity in the model architecture and training details (e.g., varying the random seed, Monte Carlo dropout\cite{yal2015mcdropout}) can be exploited to obtain the ensemble.
    
\end{enumerate}

\begin{figure}
    \centering
    \includegraphics[width=0.9\columnwidth]{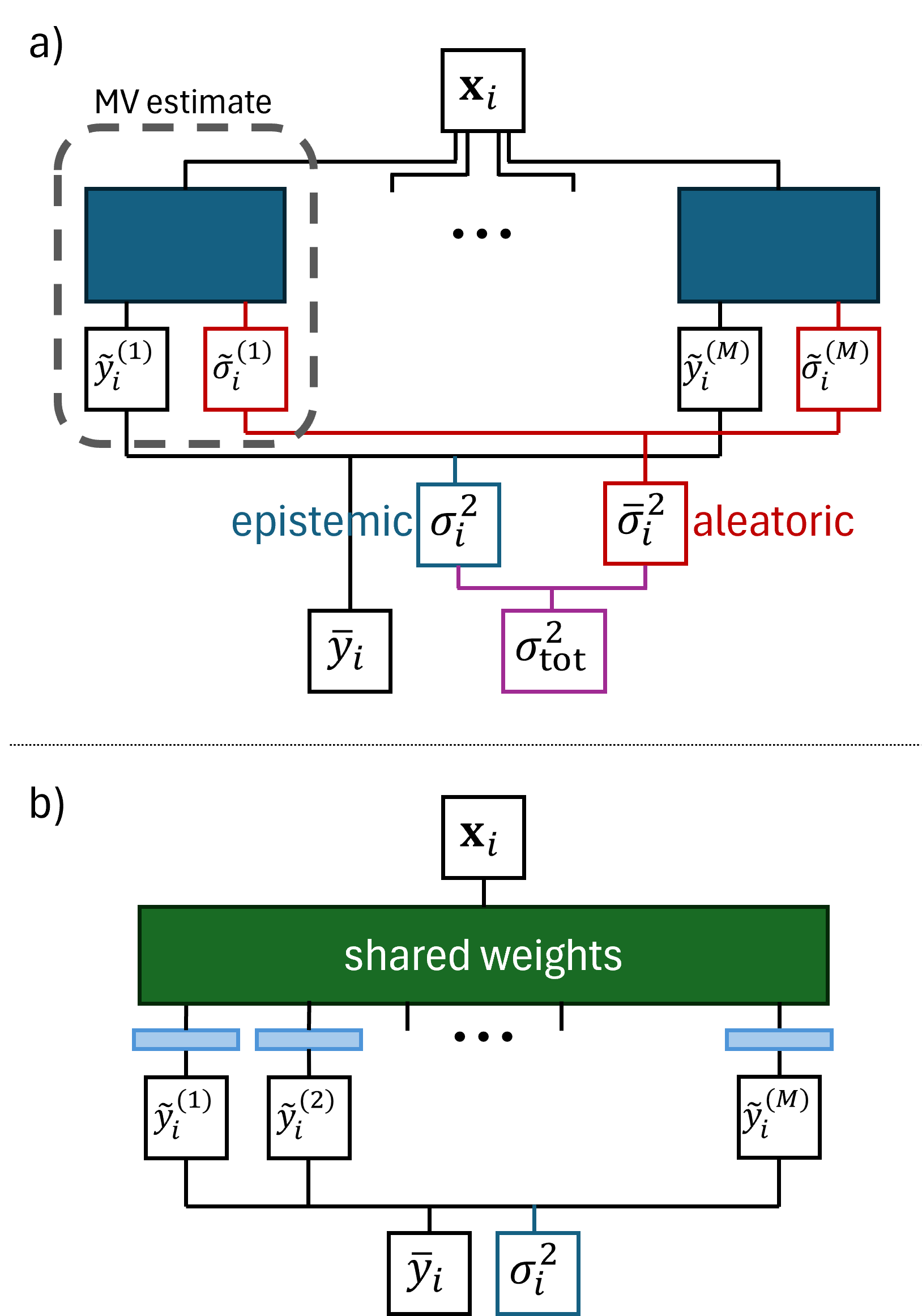}
    \caption{a) Mean variance ensemble model. b) Shallow ensemble.}
    \label{fig:MVE_shallow_ens}
\end{figure}


\subsection{Mean-variance estimation models and Mean-variance ensembles}\label{subsec: MVE est and ens}

The goal of mean variance (MV) estimation models is to predict the uncertainty $\tilde{\sigma}^2(\mathbf{x})$ affecting a given prediction $\tilde{y}(\mathbf{x})$ together with the prediction itself.
Different from ensembles, here the model is trained to directly predict the best value and its variance, rather than building an ensemble to deduce them; see also the region enclosed by the dashed line in Fig.~\ref{fig:MVE_shallow_ens}(a). 
MV estimation models are usually trained by using a negative log-likelihood loss function,\cite{nix1994estimating} that, for a single instance $\mathbf{x}_i$, reads 
\begin{equation}
\begin{split}
    \ell_i &= -\ln{p(y_i|\mathbf{x}_i,\mathbf{w})} \\
    &=\frac{1}{2}\left[ \frac{(y_i - \tilde{y}(\mathbf{x}_i))^2}{\tilde{\sigma}^2(\mathbf{x}_i)} +\ln{\tilde{\sigma}^2(\mathbf{x}_i)} +\ln{2\pi} \right]
\end{split}
\label{eq:NLL_MVestimate}
\end{equation}

Busk et al.\cite{busk2023graph} interpret $\tilde{\sigma}^2(\mathbf{x})$, predicted by MV model as an additional model output, as an \textit{aleatoric} uncertainty that may stem from random noise, data inconsistencies, or the model's inability to fit precisely, and, as such, cannot typically be reduced by collecting more data.

Ensembles models and MV estimation models can be combined to give rise to mean-variance \textit{ensembles}, introduced by  \textcite{lakshminarayanan2017simple} with the name of deep ensembles; see Fig.~\ref{fig:MVE_shallow_ens}(a). In this approach, a committee of MV estimation models is created, where each member of the committee outputs a prediction and variance pair,  $\left(\tilde{y}^{(m)}(\mathbf{x}), \tilde{\sigma}^{2,(m)}(\mathbf{x})\right)$, and then the uncertainty is estimated by summing the variance of the predictions, as in Eq. \eqref{eq:comm_var}, with the average of the variances of the committee:
\begin{equation}
\begin{split}
    \sigma^2_\mathrm{tot} (\mathbf{x}) =&  \frac{1}{M-1} \sum_{m=1}^{M} [y^{(m)}(\mathbf{x}) - \overline{y}(\mathbf{x})]^2 \\
    &+ \frac{1}{M}\sum_{m=1}^{M} \tilde{\sigma}^{2,(m)}(\mathbf{x})
\end{split}
\label{eq:MVensemble_var}
\end{equation}
which assumes that the two addenda are uncorrelated. 
Busk et al.\cite{busk2023graph} interpret the first addendum as the epistemic uncertainty and the second addendum as aleatoric uncertainty. 


\subsubsection{Shallow ensembles}

While the mean-variance ensemble approach is known to provide robust uncertainty estimates and is commonly considered the current state-of-the-art, it often suffers from the large computational cost incurred from training and evaluation of multiple models. Given that the commonly adopted ensemble size is $\geq$ 5--10, it can quickly become prohibitive for sufficiently large and complex models, especially neural networks.

In this latter case, akin to LL approximation motivated in \ref{sec:NN}, one could construct ``shallow ensembles'' (Figure \ref{fig:MVE_shallow_ens}(b)), where only the last-layer of the neural network is varied in obtaining an ensemble of models, and rest of the weights are shared across the members. Such an approach effectively mitigates the large computational cost associated with the training of multiple neural network models and their inference. ~\textcite{Kellner2024} presents a version of this approach, where a shallow ensemble of models is trained to using an NLL-like loss obtain the empirical mean and variance through Eqs.~\eqref{eq:comm_mean} and \eqref{eq:comm_var}.


\subsubsection{Mixtures of experts}

An emergent approach towards highly accurate and efficient models concerns the adoption of ensemble learning and mixture of experts (MoE) strategies. \cite{Nowlan1991}
The formulae discussed in the previous section can be easily extended to mixture-of-experts models, where the prediction for a sample $\star$ is 
\begin{equation}
    \tilde{y}(\mathbf{x}_\star) =\sum_{k=1}^K \pi_k(\mathbf{x}_\star) \, \tilde{y}^{(k)}(\mathbf{x}_\star)
\end{equation}
Here, $\pi_k(\mathbf{x})$ is an input-dependent coefficient representing the contribution of model $k$ of the mixture. The total number of models is $K$. The coefficients $\pi_k$ are normalized so that ${\sum_{k=1}^K \pi_k(\mathbf{x})}=1, \forall  \mathbf{x}$.  
Examples include soft-max assignment based on distance:\cite{Zeni2024}
\begin{equation}
    \pi_k(\mathbf{x}) = \frac{e^{s(d_k(\mathbf{x}))}}{\sum_{k'=1}^K e^{s(d_{k'}(\mathbf{x}))}}
\end{equation}
where $s(d_k(\mathbf{x}))$ labels a function of the reciprocal of the distance $d_k(\mathbf{x})$ between the point $\mathbf{x}$ and the centroid of the model dataset $k$.
Other smooth space-partitioning functions based on density have been similarly suggested:\cite{Nowlan1991}
\begin{equation}
    \pi_k(\mathbf{x}) = \frac{p^{(k)}(\mathbf{x})}{\sum_{k=1}^K p^{(k)}(\mathbf{x})}
\end{equation}
where $p^{(k)}(\mathbf{x})$ is the probability density to find $\mathbf{x}$ according to model $k$ of the mixture (e.g., in the case of Gaussian mixture models $p^{(k)}$ is the probability density function of a normal distribution defined by the $k$-th cluster's center and covariance).
By assuming that different models of the mixtures are independent among one another and characterized by the uncertainties 
\begin{equation}
    \sigma_\star^{2,(k)} = \alpha^{2,(k)} \mathbf{f}_\star^\top \mathbf{G}^{(k)}
    \mathbf{f}_\star, \qquad k=1,\ldots,K
\end{equation}
then, standard uncertainty propagation gives:
\begin{equation}
\begin{split}
    \sigma^2_\star &= \sum_{k=1}^K \left[\frac{\partial \tilde{y}_\star}{\partial \tilde{y}_\star^{(k)}}\right]^2\sigma_\star^{2,(k)} = {\sum_{k=1}^K \pi_k^2 \sigma_\star^{2,(k)}}  
\end{split}
\end{equation}

\subsection{Calibration: quantification and practicalities}\label{sec:calib}


\subsubsection{Maximum log-likelihood calibration}\label{subsec:calib_ens}

\textcite{musil2019fast} and \textcite{imbalzano2021uncertainty} have shown that the uncertainty $\sigma^2(\mathbf{x})$ estimated by the ensemble-based approach of Eq.~\eqref{eq:comm_var} can be calibrated \textit{a posteriori} by applying a global (i.e., $\mathbf{x}$-independent) scaling factor $\alpha^2$, such that $\sigma_\text{calib.}^2(\mathbf{x}) \leftarrow \alpha^2 \sigma^{2}(\mathbf{x})$.
$\alpha^2$ is chosen to maximize the log-likelihood of the predictive distribution over a validation set of $N_\mathrm{val}$ points, and is given by:
\begin{equation}
    \alpha^2 = \frac{1}{N_\mathrm{val}} \sum_{i=1}^{N_\mathrm{val}} \frac{|y_i -\bar{y}(\mathbf{x}_i)|^2}{\sigma^{2}(\mathbf{x}_i)}\label{eq:alpha2_musil}
\end{equation}
It is crucial to remark that Eq. \eqref{eq:alpha2_musil} is a biased estimator in the number of models composing the ensemble, $M$. Whenever $M$ is small, Eq. \eqref{eq:alpha2_musil} should be replaced by the bias-corrected formula
\begin{equation}
    \alpha^2 = -\frac{1}{M} + \frac{M-3}{M-1} \frac{1}{N_\mathrm{val}} \sum_{i=1}^{N_\mathrm{val}} \frac{|y_i -\bar{y}(\mathbf{x}_i)|^2}{\sigma^{2}(\mathbf{x}_i)}\label{eq:alpha2_imbalzano}
\end{equation}
from which it is also seen that at least $M=4$ members are needed.
With proper (straightforward) changes, this approach can be easily extended to the other UQ techniques outlined in the previous sections.
Notice that tracing back a clear distinction between epistemic and aleatoric components of the uncertainty, as outlined in Eq. \eqref{eq:MVensemble_var}, may be problematic after calibration.

\subsubsection{Expected vs observed uncertainty parity plots}\label{subsec:calib_bin}
Another common approach to check UQ calibration involves constructing parity plots, typically on a log-log scale, to compare the estimated variance with the observed distribution of (squared) residuals, sometimes binned according to the model’s estimated variance.\cite{lee2017deep,bigi2024prediction,Kellner2024,Levi2022,Dai2024} 
A proxy to summarize these \textit{reliability plots} is the so-called expected normalized calibration error (ENCE), recently introduced by \textcite{Levi2022}, defined by:
\begin{equation}
\begin{split}
\mathrm{ENCE} &= \frac{1}{N_\mathrm{bins}} \sum_{b=1}^{N_\mathrm{bins}} \frac{|\mathrm{RMV}(b)- \mathrm{RMSE}(b)|}{\mathrm{RMV}(b)} \\
\mathrm{RMV}(b) &\equiv \sqrt{\frac{1}{|b|} \sum_{i\in b} \sigma^2(\mathbf{x}_i)} \\
\mathrm{RMSE}(b) &\equiv \sqrt{\frac{1}{|b|} \sum_{i\in b} \left|y_i - \tilde{y}(\mathbf{x}_i)\right|^2}
\end{split}  
\end{equation}
where $|b|$ labels the number of data points in bin $b$.

If the estimated variance $\sigma^2(\mathbf{x}_i)$ follows a functional form such as Eq.~\eqref{eq:sigma2star}, the free parameter $\alpha^2$ must be adjusted, for example, following the procedure of Eq.~\eqref{eq:alpha2_musil}---to align the expected variance with the observed MSE. Notably, in log-log space, varying $\alpha^2$ results in a rigid shift of the entire plot. Thus, even for uncalibrated UQ estimates, a linear correlation between the expected and observed distributions of (squared) residuals should still be apparent. If there is a poor linear correlation, this suggests that the Gaussian-like UQ framework defined by Eq.~\eqref{eq:sigma2star} may be inadequate, as can occur in NN models with only few neurons per layer, and/or that a \textit{local} calibration $\alpha^2(\mathbf{x})$ may be necessary.

Along these lines it is worth mentioning the insightful work by \textcite{Pernot2022,Pernot2022b} which involves  stratifying the evaluation of a given calibration score, such as Eq.~\eqref{eq:alpha2_musil}, based on the predicted uncertainty: the dataset is split into \textit{subsets} where the predicted 
uncertainty falls within certain ranges (e.g., low, medium, or high uncertainty predictions); calibration metrics are then calculated independently for each subset. This allows for a more detailed analysis of how well the model is calibrated across different levels of predicted uncertainty.
By performing different types of stratification/binning of the dataset, \textcite{Pernot2023} shows that it is also possible to distinguish between
\textit{consistency} (the conditional calibration with respect to prediction uncertainty) and \textit{adaptivity} (the conditional calibration with
respect to input features), and that good consistency and good adaptivity are rather distinct and complementary calibration targets.

\subsubsection{Miscalibration area}
In the literature,\cite{Tran2020} (mis)calibration is often discussed in terms of the similarity between the expected and observed cumulative distribution functions (CDFs) of residuals. A model is considered calibrated when the miscalibration area between these CDFs is small, indicating consistency between the predicted and actual uncertainties. The sign of the miscalibration area can further reveal whether the model's estimated uncertainties are under- or over-confident, providing additional diagnostic insight. 

Nonetheless, methods based on the miscalibration area can be challenging to interpret. In fact, they provide an indirect assessment of UQ quality, since, by comparing CDFs, they inherently compare higher moments of the distributions. For example, two distributions may share similar second moments--quantities typically interpreted as uncertainty---but diverge significantly in their CDFs due to deviations from Gaussianity, such as skewness or heavy tails. This conflation of uncertainty with other aspects of distribution's shape highlights a key limitation of such approaches. 

For this reason, we recommend prioritizing the calibration strategies discussed in Secs. \ref{subsec:calib_ens} and \ref{subsec:calib_bin}, which more clearly align with the intended interpretation of uncertainty.  

\subsection{Conformal prediction}

Most of the UQ strategies described so far employ input-dependent uncertainties whose evaluation are largely independent of the values of the targets in the dataset (see point 3.~in Sec.~\ref{sec:direct_estimates}). 
A complementary and rather opposite idea is based on conformal predictions (CPs), which---for regression tasks---provide a way to construct  \textit{confidence intervals} for a continuous target prediction $\tilde{y}_\star$, such that the intervals contain the true value with a predefined probability (e.g., 95\%).
The idea of CPs stems from seminal concepts developed by Ronald Fisher in the 1930s,\cite{fisher1935fiducial} and then applied to the context of ML by Vladimir Vovk and collaborators in the 1990s (for a pedagogical review, see e.g.~\textcite{shafer2008tutorial}).
In a nutshell, the CP procedure reads as follows:
\begin{itemize}
    \item train a regression model to give predictions $\tilde{y}_i$
    \item use a separate calibration dataset $\mathcal{C}$
    \item compute the \textit{nonconformity score} $s_{i\in\mathcal{C}}$ (typically the absolute error, $s_i = |y_i - \tilde{y}_i|$)
    \item determine the $(1-\alpha)$-quantile $q_{1-\alpha}$ of the sorted scores. For instance, if $\alpha=0.05$, then 95\% of the scores $s_{i\in \mathcal{C}}$ shall lie below the value $q_{0.95}$
    \item for a new input $\mathbf{x}_\star$ construct the prediction interval as $[\tilde{y}_\star - q_{1-\alpha},\tilde{y}_\star + q_{1-\alpha}]$. 
\end{itemize}
The prediction interval contains the true $y$ with at least $1-\alpha$ confidence.
The key assumption in CP is that the new (test set's) error distribution is representative of both the training and calibration sets error distribution. This allows $
1-\alpha$, derived from the calibration data $\mathcal{C}$, to apply universally across all inputs $\mathbf{x}$.
As a result, $q_{1-\alpha}$ acts as a \textit{global} threshold for constructing prediction intervals, a measure of the model's ``typical error'' that encapsulates how large the prediction errors tend to be (up to the $
(1-\alpha)$-quantile), independent of any specific input $\mathbf{x}$.
Nonetheless, if the new set of inputs is significantly out of distribution, meaning it differs substantially from the training and calibration data, the assumptions underlying CP may no longer hold, and the prediction intervals obtained from CP might lose their validity.

In the context of atomistic modeling of materials, CPs have been recently used, e.g., in \textcite{Hu2022}, for the UQ of ML interatomic potentials.

\subsection{Are all uncertainty estimates the same?}

Standardized benchmarks and evaluation protocols for ML model accuracies in atomistic modeling have been established only recently. Shortly after, limitations of these benchmarks where also identified, and this continues to be an area of ongoing research.
Similarly, a consensus on UQ methods reliability is still lacking, since no unique set benchmarks for uncertainty estimation has been developed yet.

\textcite{Tran2020} compared the accuracy and uncertainty of multiple machine learning approaches for predicting the adsorption energy of small molecules on metals. The most effective approach combines a convolutional neural network (CNN) for feature extraction with a Gaussian process regressor (GPR) for making predictions. This hybrid model not only provided accurate adsorption energy estimates but also delivers reliable uncertainty quantification. 

\textcite{Tan2023} evaluated ensembling-based uncertainty quantification methods against single-model strategies, including mean-variance estimation, deep evidential regression, and Gaussian mixture models. Results, using datasets spanning in-domain interpolation (rMD17) to out-of-domain extrapolation (bulk silica glass), showed that no single method consistently outperforms the others. Ensembles excel at generalization and robustness, while MVE performs well in-domain, and GMM is better suited for out-of-domain tasks. The authors concluded that, overall, single-model approaches remain less reliable than ensembles for UQ in NNIPs. 

\textcite{Kahle2022} reported that  NN potentials ensembles may result overconfident, underestimating the uncertainty of the model. Further, they require to be calibrated for each system and architecture. This was verified across predictions for energy and forces in an atomic dimer, an aluminum surface slab,
bulk liquid water, and a benzene molecule in vacuum. Bayesian NN potentials, obtained by sampling the posterior distribution of the model parameters using Monte Carlo techniques, were proposed as an alternative solution towards better uncertainty estimates.


Further, the integration of UQ methods with existing machine learning architecture is often streamlined for one specific approach only (\cite{Litman2024, Moriarty2022}) and it is rarely the case that one single workflow allows for the adoption of a diverse set of ML UQ methods.

\subsection{Size extensivity of uncertainty estimates}

Another important consideration in ML for atomistic modeling is the size extensivity of properties targeted by the ML models, and how that propagates to the uncertainty estimates. Take ML interatomic potentials as an example, where the models are trained to predict total energies of chemical systems as a sum of atomic contributions. It is unclear how the uncertainties of the systems grow with their size. One could rationalize two extrema: one is a perfectly crystalline system with all-equivalent atomic environments, leading to maximal correlation between local predictions and hence uncertainties would strictly grow as $N$, the number of atoms. The other would be a dilute gas of atoms or molecules with no correlation, leading to the growth of uncertainty in quadrature, i.e., scaling with $\sqrt N$. Real chemical systems are expected to have components that can be distinguished with both scaling behaviors.

Kellner and Ceriotti\cite{Kellner2024} have investigated the size extensivity of uncertainty estimates for bulk water systems using their shallow ensembling method for a deep NN model. In their analysis, they decomposed the uncertainties on differently sized bulk water systems into ``bias'' and ``residual'' terms. The bias term, computed by taking the absolute difference between the mean predicted and reference energies for a given system size, was found to scale roughly with $N$. The remaining residual term, which would largely capture the random distortions of the water molecules, was then found to correlated with $\sqrt N$. Given the significant contributions from both terms, their experiments showcase the non-trivial trends in the size extensivity of ML model uncertainties for real material systems, which exposes the limitations of approaches where extensivity is ignored or a naive scaling law is assumed.

\subsection{Uncertainty propagation}

Besides being an alternative strategy with respect to the direct application of the formulae of Sec.~\ref{sec:direct_estimates}, the use of ensembles is particularly useful in several physical applications that require the propagation of uncertainty to derived quantities that are function $f$ of the regressor's output, $z(\mathbf{x}) = f(y(\mathbf{x}))$. 
In fact, only in few simple cases uncertainty can be propagated analytically from the UQ formulae presented in Sec.~\ref{sec:direct_estimates} in the form of Eq.~\eqref{eq:sigma2star}, as it is the case, for instance,  whenever the linear approximation
\begin{equation}
    \sigma_z(\mathbf{x}_\star) \approx \left.\frac{\mathrm{d}f}{\mathrm{d}y}\right|_{\mathbf{x}_\star} \, \sigma_\star
\end{equation}
is sufficiently accurate.
In the general scenario one can  easily resort to \textit{explicit sampling} of the models' distribution. 
\begin{equation}
\begin{split}
    &z^{(m)}(\mathbf{x}) = f\left(y^{m}(\mathbf{x})\right)\\
    &\overline{z}(\mathbf{x}) = \frac{1}{M}  \sum_{m=1}^{M} z^{(m)}(\mathbf{x})\\
    &\sigma_z^2(\mathbf{x}) = \frac{1}{M-1} \sum_{m=1}^{M} [z^{(m)}(\mathbf{x}) - \overline{z}(\mathbf{x})]^2
\end{split}
\end{equation}


In ML-driven atomistic simulations, UQ is also needed to single out the uncertainty ascribable to ML models from the statistical one due to a poor sampling (i.e. too short trajectories): in fact it would be pointless to run very long MD simulations if the uncertainty due to ML cannot be lowered below a given threshold. \\

A more subtle problem arises in the realm of MLIPs, whenever one aims at propagating uncertainty to thermostatic observables (e.g., the radial distribution function, the mean energy of a system, etc.) where ML uncertainty on the energy of a given structure enters the Boltzmann weight of thermodynamic averages, or -- equivalently under the erdogic hypothesis -- affects the sampling of the phase space via molecular dynamics simulations.  
In ~\textcite{imbalzano2021uncertainty}, the availability of model-dependent predictions was leveraged to propagate uncertainty to thermostatic observables while running a single trajectory with the mean MLIP of the ensemble, by applying simple reweighing strategies. 

For instance, consider the case of training a committee of $M$ ML interatomic potentials to learn the ML potential energy surface $V(\mathbf{r})$, where $\mathbf{r}$ indicates the set of positions of all the atoms of a system. The potential energy of the $i$-th model is labeled by $V^{(i)}(\mathbf{r})$, and the mean potential energy of the committee by $\overline{V}(\mathbf{r})$, as in Eq.~\eqref{eq:comm_mean}. 
Then, for a given observable $a(\mathbf{r})$ of the atomic positions, its canonical average, computed by sampling the configurational space according to the Boltzmann factor $\exp[-\beta V^{(i)}(\mathbf{r})]$, where $\beta=(k_\mathrm{B}T)^{-1}$, can be equivalently expressed in terms of a canonical average using the Boltzmann factor $\exp[- \beta \overline{V}(\mathbf{r})]$ associated to the mean potential of the committee as:
\begin{equation}
    \langle a\rangle_{V^{(i)}} = \frac{\int \mathrm{d}\mathbf{r} \, w^{(i)}(\mathbf{r}) \, a(\mathbf{r}) \, e^{-\beta\overline{V}(\mathbf{r})} }{\int \mathrm{d}\mathbf{r} \, w^{(i)}(\mathbf{r}) \, e^{-\beta \overline{V}(\mathbf{r})} }\label{eq:reweighting}
\end{equation}
where $w^{(i)} \equiv \exp[-\beta(V^{(i)}(\mathbf{r}) - \overline{V}(\mathbf{r}))] $. 
Therefore, by performing a single experiment to sample the configuration space (e.g., via Monte Carlo or molecular dynamics) using the mean potential of the committee, one can post-process the result to obtain the set $\langle a\rangle_{V^{(i)}}$,  $i=1,\ldots,M$, whose standard deviation across the committee quantifies the uncertainty on the thermodynamic average $\langle a\rangle$. Statistically more stable approximations also exist, based on a cumulant expansion, to overcome sampling efficiency issues stemming from direct application of Eq.~\eqref{eq:reweighting}.\cite{ceriotti_efficient_2012,imbalzano2021uncertainty,Kellner2024}

Looking ahead, fundamental questions remain. E.g., from a physical perspective:
{\it what are the key ingredients towards the definition of a rigorous theory for uncertainty propagation for time-dependent thermodynamic observables, such as correlation functions?}
Such a question is relevant for spectra and transport coefficients, obtained from ML-driven molecular dynamics simulations,\cite{sosso_thermal_2012, verdi_thermal_2021, tisi_heat_2021, pegolo_temperature_2022, malosso_viscosity_2022, langer_heat_2023, tisi_thermal_2024, pegolo_thermal_2024, pegolo_transport_2025, drigo_seebeck_2024} since its answer would make it possible to quantify the uncertainty on these dynamical observables in an efficient way, bypassing time-consuming, brute-force approaches that require running several trajectories.

\section{Uncertainty and robustness}

Having reviewed a wide range of UQ methods for ML models, we now extend our discussion to the ``robustness'' of the models and their predictions. By robustness, we refer to the ability of a model to maintain good accuracy and precision under various types of perturbations, noise, and adversarial conditions in the provided input. Approaching the robustness of ML models requires the knowledge of when and where the ML model fails or stops being applicable, \textit{even in the absence of target values} unlike the case of UQ. Through such an understanding and quantification of ML model robustness, one can propose efficient methods for rational dataset construction/augmentation and active  learning for ML training.

In the context of atomistic modeling, robustness is important for novel materials discovery, where models are often used to predict the properties of new phases or materials that lie outside the existing dataset. 
The concept of robustness can also be straightforwardly extended to the predictions of ``local'' (e.g., atom-centered) or ``component-wise'' (e.g., range-separated) quantities of the chemical systems, which do not correspond to physically observable targets. This is especially relevant for ML models constructed to make global predictions on the system as the sum of local and component-wise predictions on distinguishable parts and their associated features. This is indeed a common practice in ML for atomistic modeling.


\subsection{A geometrical perspective on in- and out- of distribution}

Intuitively, a prediction is likely to be accurate and precise if it takes place in the region corresponding to the distribution of training points. 
Towards the definition of robust prediction, it is then relevant to explore the definition of in- and out-of distribution.
A first perspective to this end concerns a geometric framework and a convex hull construction.
The convex hull of a set of training samples $\mathcal{X} = \{\mathbf{x}_1, \mathbf{x}_2, \ldots, \mathbf{x}_n\}$ in the feature space is defined as the smallest convex set that contains all points in $\mathcal{X}$. Mathematically, the convex hull is expressed as:

\[
\text{Conv}(\mathcal{X}) = \left\{\mathbf{x} \; \middle| \; \mathbf{x} = \sum_{i=1}^n \alpha_i \mathbf{x}_i, \; \alpha_i \geq 0, \; \sum_{i=1}^n \alpha_i = 1 \right\},
\]

where $\alpha_i$ are convex coefficients ensuring that any point $\mathbf{x}$ inside the hull is a weighted combination of the training samples $\mathbf{x}_i$.

\begin{figure}
    \centering
    \includegraphics[width=\columnwidth]{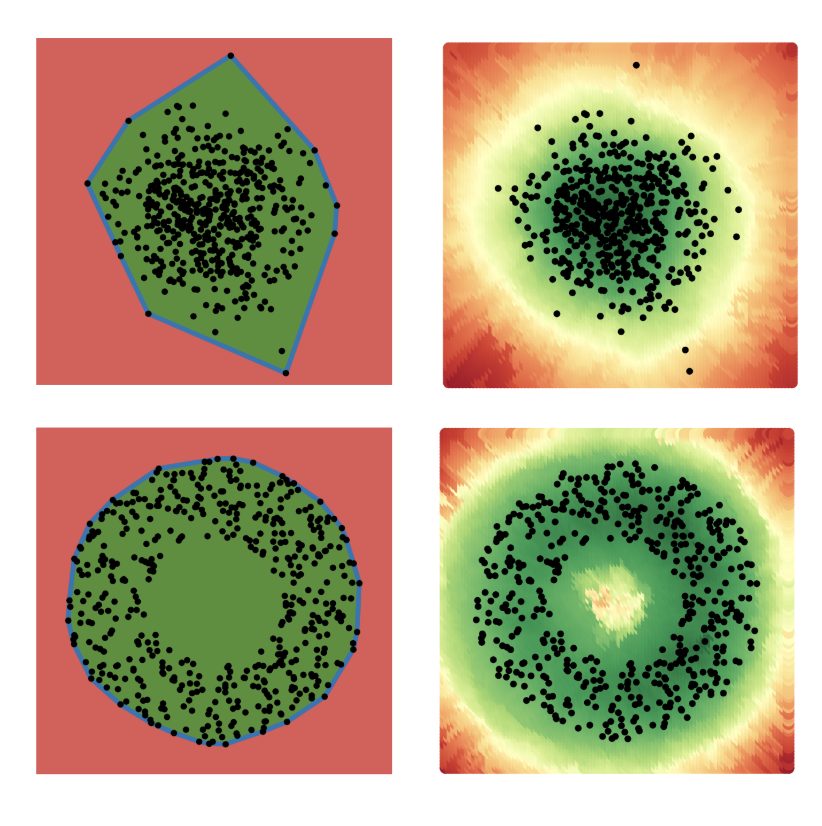}
    \caption{The four panels illustrate two example distribution of points in two dimensions. In the left panels, in- and out- of distribution regions are categorized according to a convex-hull geometric construction (left). In the right panels in- and out- of distribution are defined according to a density-based criterion. For the left panels, the green region is defined as in-distribution, the red region is out-of-distribution. For the right panels, color from green to red highlight areas moving from in- to out- of distribution. Figure courtesy of Claudio Zeni.}
    \label{fig:coverage}
\end{figure}

This construction provides a method to distinguish in-distribution samples, which lie within $\text{Conv}(\mathcal{X})$, from out-of-distribution samples, which fall outside the convex hull. Extrapolation, in this context, refers to the model's attempt to make predictions for such out-of-distribution points by extending patterns learned from the training data, often resulting in increased uncertainty and reduced accuracy  (Figure \ref{fig:coverage} left panels).

The convex hull evaluation faces significant challenges in high-dimensional spaces. 
The computational cost of constructing and evaluating convex hulls increases with dimensionality, making this approach impractical for large-scale, high-dimensional machine learning tasks. 
Even more importantly, the number of points required to approximate the convex hull grows exponentially with the dimensionality of the feature space, a problem commonly referred to as the ``curse of dimensionality.'' Consequently, the convex hull becomes increasingly sparse in high dimensions, causing most points in the space to be classified as out-of-distribution.\cite{balestriero2021}

Importantly, while the intrinsic dimensionality of high-dimensional representation may be still low, low-dimensional projections (e.g., D=2 or D=3) for visualization or analysis, can introduce artifacts that misrepresent the true relationships in the data, such as incorrectly classifying in-distribution samples as out-of-distribution due to oversimplified boundaries. 
This is also relevant in machine learning for atomistic modeling, where the information high-dimensional representation can be reduced 
to a small but not too small amount of principal components. \cite{Zeni2022}

\subsection{A statistical perspective on in- and out- of distribution}

An alternative to the convex hull for defining in- and out-of-distribution samples is to use a density-based method. Here one evaluates the likelihood of a sample belonging to the training distribution by estimating the sampling density in the feature space (Figure \ref{fig:coverage} right panels) 

In an adaptive \(k\)-nearest-neighbor (k-NN) density estimation procedure proposed by Zeni \emph{et al.}~\cite{Zeni2022}, each test point \(\mathbf{x}^*\) is temporarily inserted into the training set so that its \(k^*\) nearest neighbors among the training samples can be identified. This process makes it possible to compute the local density at \(\mathbf{x}^*\) via:
\begin{equation}
    \rho(\mathbf{x}^*) = \frac{k^* - 1}{M \, V^*},
\end{equation}
where \(M\) is the total number of training examples, and \(V^*\) is the volume corresponding to the \(k^*\) neighbors . The number \(k^*\) is selected in an adaptive manner for each test point to optimize the precision of the resulting density estimate. Moreover, the volume \(V^*\) is determined by the hypersphere of dimension \(d\), where \(d\) represents the intrinsic dimensionality of the training set as computed using the TwoNN estimator \cite{Facco2017, Glielmo2022}.

Through this methodology, the resulting metric reflects the degree to which an unseen atomic environment lies in a well-sampled portion of the representation space. Importantly, it also correlates with the errors observed in machine-learned regression potentials. 
Furthermore, the same authors report a strong consistency between this density-based measure and a model-specific error estimator, namely the predictive uncertainty from a committee of models trained on different subsamples of a larger training set.

In related work, \textcite{schultz2024determiningdomainmachinelearning} utilized kernel density estimation in feature space to evaluate whether new test data points fall within the same domain as the training samples. Their approach illustrates that chemical groups traditionally considered unrelated exhibit pronounced divergence according to this similarity metric. Moreover, they show that higher dissimilarity correlates with inferior predictive performance (manifested as larger residuals) and less reliable uncertainty estimates. They additionally propose automated methods to define thresholds for acceptable dissimilarity, enabling practitioners to distinguish between in-domain predictions and those lying outside the model's scope of applicability.

We similarly consider the work of \textcite{Zhu2023} in the context of statistical methods to define extrapolation and interpolation and in- and out- of distribution.
Given a specific training set, a feature vector for each point is derived from the latent space representation of a NequIP \cite{batzner2022} model. Next, a Gaussian mixture model (GMM) is fitted on this distribution. A negative log-likelihood can be then obtained by evaluating the fitted GMM on the feature vector associated to any test point.
Higher negative log-likelihood were observed for points resulting in higher predictions uncertainty. 

To conclude, we note that, while statistical estimates of in- and out- of distribution are of interest because of their efficiency and effectiveness, questions remains. The magnitude of these metrics depends on the chosen representation,\cite{Zeni2022} and its precise correlation with the mean absolute error (MAE) is contingent upon both the system and the model employed.
%


\subsection{Transferability}
\begin{figure}
    \centering
    \includegraphics[width=1\linewidth]{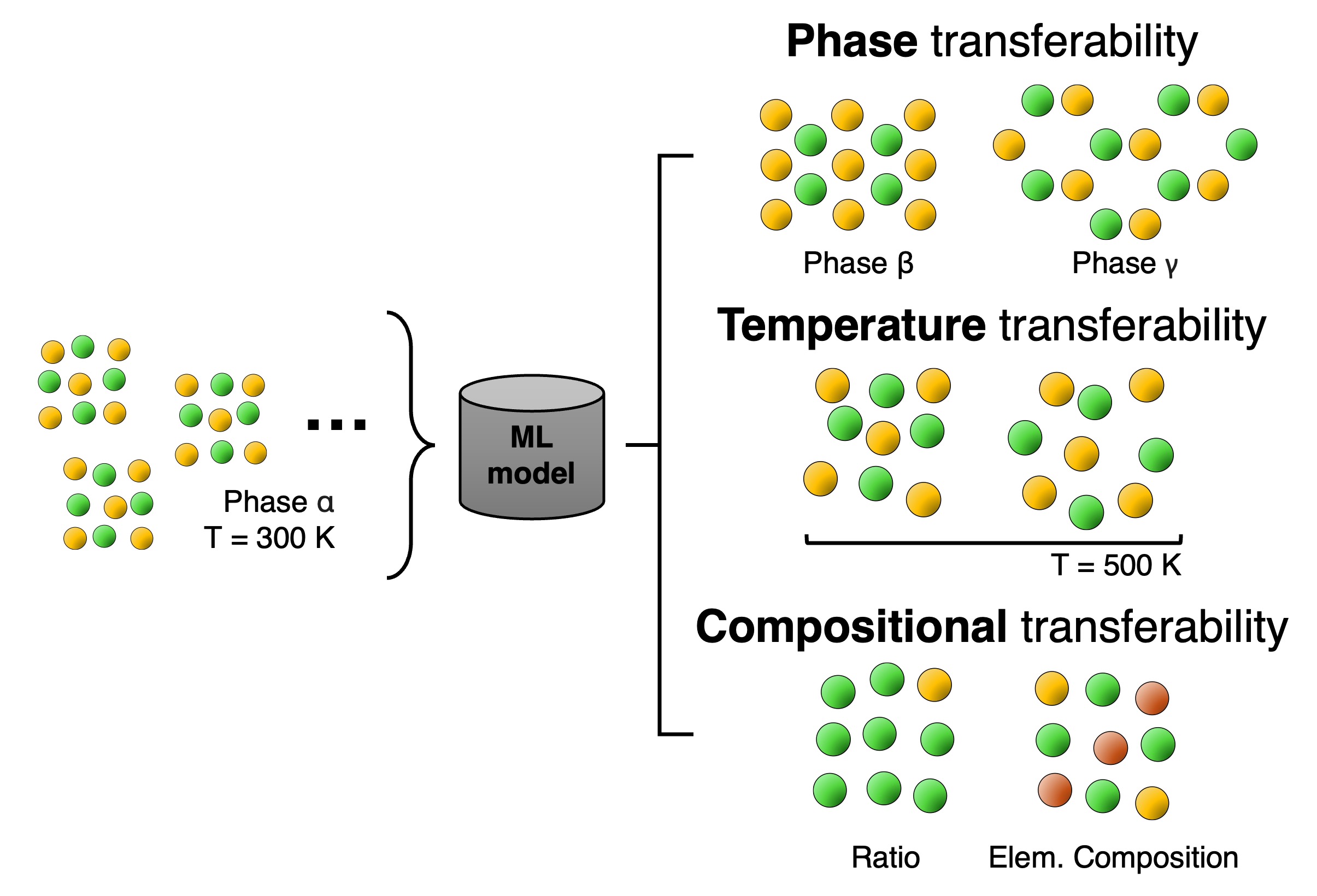}
    \caption{Illustration of possible transferability tests. A model is trained on initial database (left) consisting of structures in phase $\alpha$ sampled at $T=300$ K. Its transferability may be tested for the case of different phases ($\beta$ and $\gamma$ in the illustration), temperatures (e.g., $T=500$ K), or compositions (i.e. different stoichiometries or elemental compositions). }
    \label{fig:transferability}
\end{figure}

Transferability, in the context of machine learning for atomistic modeling, is often defined as the ability of a ML model to maintain its accuracy when applied to structures sampled under conditions different from those in the training dataset. However, the definition of these "different conditions" has remained somewhat weak, and can be summarized as follows (also illustrated in \ref{fig:transferability}):
\begin{itemize}
    \item \emph{Phase Transferability} refers to the ability of an ML potential trained on certain phases of a material to accurately predict properties of other ones (e.g. different polymorphs or phases), assuming both are sampled at similar temperatures.\cite{Kandy2023,GuidarelliMattioli2023, Wellawatte2023}
    \item \emph{Temperature Transferability} concerns the accuracy of the ML potential when, e.g., trained on structure sampled at high temperatures and tested on structures at lower temperatures, or viceversa. \cite{Zeni2018,Owen2024, GuidarelliMattioli2023, Stocker2022}
    \item \emph{Compositional Transferability} refers to the ability of a ML model when providing predictions for systems with unseen compositions with accuracy comparable to known stoichiometries. This can refer both to predictions for unseen stoichiometries, or for unseen elements (alchemical learning).\cite{Imbalzano2021, Lopanitsyna2023, nam2024interpolationdifferentiationalchemicaldegrees}
\end{itemize}

The standard according to which a model is deemed transferable across different conditions remains rather flexible too.
Criteria used to relate transferability and model accuracy so far have included:
\begin{itemize}
    \item The error incurred by the model in the test domain is \emph{comparable} with the one observed for the training domain.
    \item The error in the test domain is \emph{sizably larger} from the one in the training domain, but remains \emph{acceptable} for practical purposes (e.g., energy errors below 10 meV/atom, force errors below 100 meV/\AA).
    \item Simulations remain \emph{stable over long timescales}, showing no significant energy drift or sampling of unphysical configurations.
\end{itemize}
The lack of a rigorous and standardized definition of transferability challenges our attempt to unify conclusions drawn from studies concerned in assessing ML model transferability in the context of atomistic modeling.
Heuristic observations on transferability report that:
\begin{itemize}
    \item \emph{Phase Transferability}: There is often a trade-off between accuracy and generality when applying ML potentials across different phases. This trade-off is generally acceptable for many practical applications.\cite{Zeni2018, Zeni2024}
    \item \emph{Temperature Transferability}: Models trained on high-temperature data tend to generalize well to lower-temperature conditions. \cite{GuidarelliMattioli2023,Owen2024}
    \item \emph{Compositional Transferability}: Interpolation within the compositional space is generally feasible, but extrapolation to entirely new stoichiometries or elements (e.g., alchemical learning) poses significant challenges, unless tailored schemes are adopted.\cite{Imbalzano2021, Lopanitsyna2023}
\end{itemize}

The relationship between a model's complexity and its transferability has also been a subject of discussion. In principle, more flexible models are more susceptible to overfitting, which can reduce transferability. Empirically, this tendency has been observed in ML methods based on feed-forward neural networks.\cite{Kandy2023}
Importantly, modern high-order graph convolutions and/or physics-inspired (e.g., symmetry conserving) architectures have not exhibited this limitation, suggesting that increased complexity does not necessarily compromise transferability, at least within the data- and parameter-sizes considered in those applications.\cite{batatia2024}

\subsection{Quantifying Robustness}\label{sec:pred_rigidity}

To meaningfully interpret the robustness of a prediction in itself, study its dependence on the datatset composition, and compare the robustness of a prediction on one input with another, the necessity to perform such an assessment in a quantitative manner arises.

To address this problem, recently, Chong and coworkers have introduced the concept of {\it ``prediction rigidities''} (PRs),\cite{chong2023robustness,bigi2024prediction,chong2024rigidityfd} which are metrics that quantify the robustness of ML model predictions. Derivation fo the PRs begin from considering the response of ML models to perturbations in their predictions, via their loss, by adopting the Lagrangian formalism. A modified loss function can be defined as shown below:
\newcommand{\oL}{\mathcal{L}}
\newcommand{\nL}{\hat{\mathcal{L}}}
\newcommand{\Ho}{\mathbf{H}_o}
\newcommand{\oow}{\mathbf{w}_o}
\newcommand{\ooL}{\mathcal{L}_o}
\newcommand{\now}{\hat{\mathbf{w}}_o}
\newcommand{\noL}{\hat{\mathcal{L}}_o}
\begin{equation}
    \nL(\mathbf{w}) = \oL (\mathbf{w}) + \lambda \left[\epsilon_\star - (\bm{\phi}^o_\star)^{\sf T} (\mathbf{w} - \oow) \right] 
\end{equation}
where $ \bm{\phi}^o_i \equiv \left.\frac{\partial \tilde{y}_\star}{\partial \mathbf{w}}\right|_{\oow}$, and $\epsilon_\star$ is the perturbation of the model prediction for the input of interest denoted by $\star$. It is then possible to perform constrained minimization of this new loss, leading to the following expression that solely depends on $\epsilon_\star$:
\begin{equation}
    \noL(\epsilon_\star) = \ooL + \frac{1}{2}\,\left.\frac{\partial^2 \noL}{\partial \epsilon_\star^2}\right|_{\epsilon_\star = 0} \epsilon_\star^2
    \label{eq:prloss}
\end{equation}
Here, the ``curvature'' at which the model responds to the perturbation in the prediction is given by ${\partial^2 \noL}/{\partial \epsilon_\star^2}$, which can be further derived as follows: 
\begin{equation}
    \left.\frac{\partial^2 \noL}{\partial \epsilon_\star^2}\right|_{\epsilon_\star =0} \equiv \frac{1}{(\bm{\phi}^o_\star)^{\sf T} (\Ho)^{-1}\bm{\phi}^o_\star}
\end{equation}
One can recognize the crucial connection between the $\Ho$ appearing in this expression and the $\Ho$ defined in Eq. \eqref{eq:Laplace_approx}, as well as the orignal Eq. \eqref{eq:sigma2star} defined for the Mahalanobis distance. Note that $\Ho$ is often approximated as the sum of the outer products of \emph{structural} features over the training set, which would be the sum or mean of atomic features of the given structure and an indirect approach to consider the ``groupings'' of local environments that are present as structures in the training set.

A few important remarks should be made here:
\begin{itemize}
    \item  absence of any calibration parameters in the PRs hint that these are purely dataset and representation-dependent parameters, and hence distinct from being a UQ metric; 
    \item dependence on the dataset and model training details is determined by $\Ho$, which can adopt a Gauss-Newton approximation scheme and be computed in a similar manner as Eq. \eqref{eq:Gauss-Newton_Hessian}, and remains constant for a given model;
    \item there is freedom in how $\star$ is defined --- it is possible to compute the PRs for any data point as long as it is definable within the input parameter space, furthermore, one can also target specific local predictions or component-wise predictions of the model, resulting in \emph{local} PR (LPR) or \emph{component-wise} (CPR) that quantitatively assess the robustness of intermediate model predictions that do not have corresponding physical observables.
\end{itemize}

The robustness metrics introduced thus far are solely dependent on the dataset distribution (i.e. structural diversity of material systems and their local environments) and remain detached from the distribution of the target metric. One must be mindful of the repercussions, which is that complexity of the target quantity landscape is ignored: if the target landscape is smooth, learning may require fewer data points to achieve the target accuracy; if it is rough, more data points would be needed to resolve the complex landscape and achieve desirable accuracy.

The quality of data and representation may be similarly relevant.
For example, as 
discussed by Aldeghi et al.\cite{aldeghi2022roughness}and van Tilborg et al.\cite{vantilborg2022activitycliffs}, 
{\it ``activity cliffs''}---instances where structurally similar pairs of molecules that exhibit large differences in the targets---negatively affect the model performance.
By learning a representation, e.g., through contrastive learning, that correctly separates such structures the learning problem is simplified.
Also, a modified Shapley analysis was also proposed for analysing and interpreting the impact of a datapoint in the training set on model prediction outcomes \cite{Liu2023, Barnard2023}.

\subsection{Dataset improvement and active learning}

In atomistic modeling, tasks such as identifying global minima in complex energy landscapes and estimating statistical observables from molecular dynamics sampling require efficient exploration of vast and high-dimensional spaces. 
A recurring question then emerges: {\it What (additional) data should one select to gather, to build a ``better'' model (i.e., capable of more reliable/robust predictions)?}
The problem of optimal data selection is in fact crucial in two main scenarios:
\begin{enumerate}
    \item Generation of new targets is relatively expensive and/or time consuming. In such a case one may like to know \textit{in advance} for which new data point inputs $\mathbf{x}$ compute the target $y$ (i.e. assign a label);
    \item There is a vast pool of data and one aims to select a subset of data points. This is critical for, e.g., the construction of the representative set of sparse kernel models, although it is common in modern deep learning training strategies to use all the data at disposal.
\end{enumerate}

A first example of workflows iteratively improving the accuracy of an atomistic model was the ``learn on the fly'' hybrid scheme proposed by \textcite{Csanyi2004} Here, fitted potentials (based upon an analytical formulation \cite{Csanyi2004} or machine learning \cite{Li2015}) are refined using a predictor-corrector algorithm and quantum calculations to ensure reliable simulations of complex molecular dynamics.
%
%
%

Since the units of the uncertainty naturally allow for the definition of interpretable thresholds and tolerance criteria, uncertainty can be naturally adopted as the metric to identify configurations where model predictions are too uncertain, for which additional information is necessary to steer the model towards more robust predictions.

Numerous active learning schemes for interatomic potentials based upon uncertainty thresholds have been proposed in the last years, encompassing a variety of materials and chemistry, from heterogeneous catalysis \cite{Vandermause2022, Tran2018} to energy materials, \cite{Marcolongo2020} from reactions in solutions \cite{Rossi2020, Zhang2024} to molecular crystals \cite{Anelli2024}.
More recently, biasing the sampling towards configurations corresponding to highly uncertain prediction was brought forward as a strategy to ensure the collection of varied training set, and the training of a model presenting an uncertainty always below a specific threshold across a (large) region of interest in the configurational space.
\cite{Schran2020,vanderOord2023, Kulichenko2023, Schwalbe-Koda2021, Zaverkin2024}

\begin{figure}
    \includegraphics[width=0.45\textwidth]{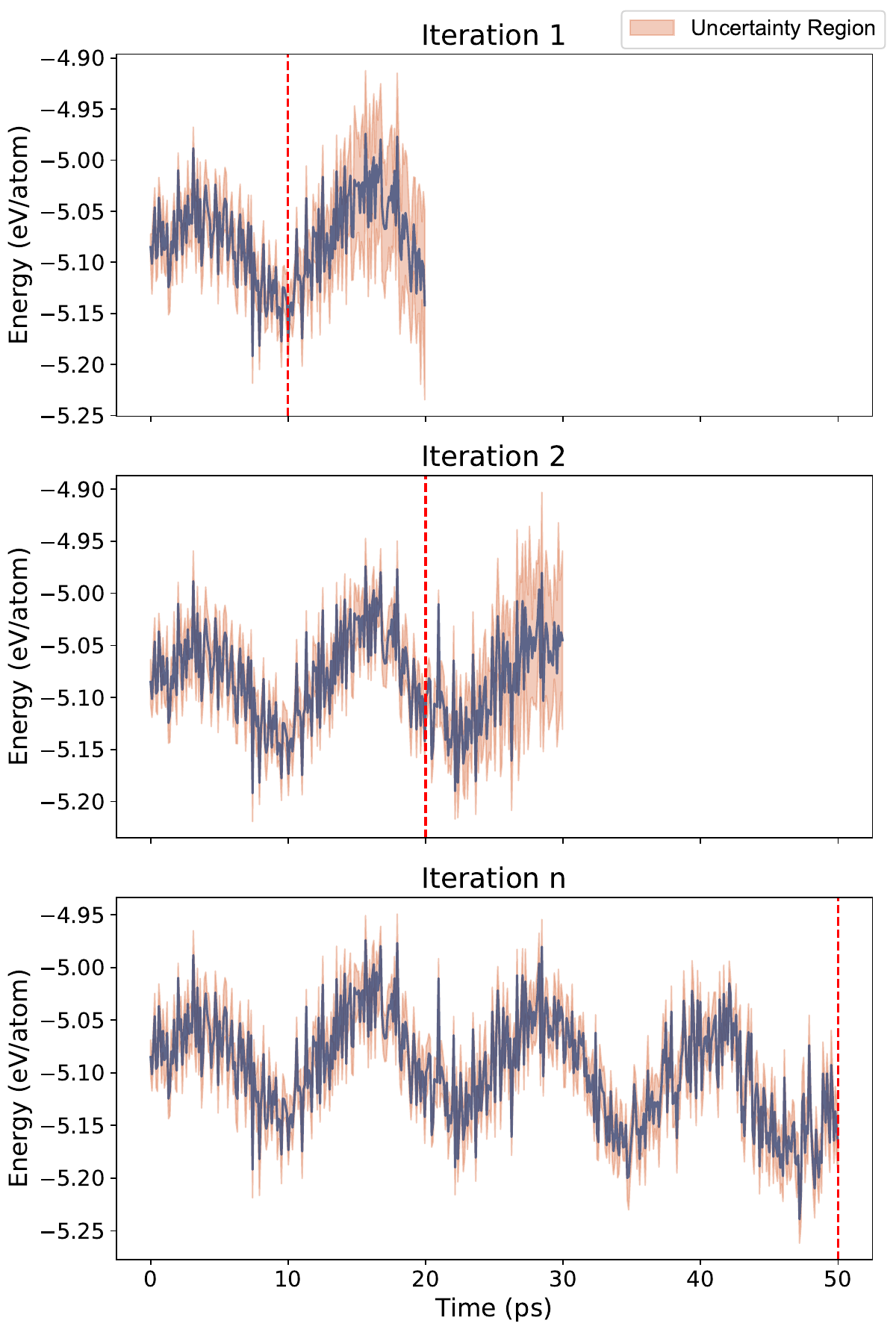}
    \caption{Example of active learning for energy evolution over time for different iterations. The orange region indicates uncertainty. In Iteration 1 and Iteration 2, uncertainty increases significantly after the red dashed line leading to termination of the simulation. The last iteration (Iteration n) represents the case where the previous active learning iteration result in a stable, accurate, and precise simulation.} 
    \label{fig:actilearn}
\end{figure}

In the next subsections,
we show that several ``active learning'' approaches to sequentially select new, optimal data points can be framed in the context of the maximum gain of information, as first discussed by \textcite{mackay1992information} in the context of the Bayesian interpretation of learning.
We also show that apparently model-free approaches do effectively identify new points where uncertainties would be the largest.

\subsubsection{Maximizing information gain}\label{subsec:information_gain}

Consider a dataset $\mathcal{D}$ of $N_\mathrm{train}$ points with feature matrix $\mathbf{F}$.
From the information theory standpoint, we can define the Shannon entropy
\begin{equation}
    S \equiv -\int \mathrm{d}\mathbf{w} \, p(\mathbf{w}|\mathcal{D}) \, \log[ p(\mathbf{w}|\mathcal{D})]
\end{equation}
where $p(\mathbf{w}|\mathcal{D})$ is a probability measure of the parameters, the weights $\mathbf{w}$, given the model architecture and dataset $\mathcal{D}$. In the Bayesian interpretation, $p(\mathbf{w}|\mathcal{D})$ is the posterior distribution of the weights, which assumes the form of a multivariate Gaussian distribution, Eq.~\eqref{eq:Laplace_approx}, whenever the model is linear or a Laplace approximation around the MAP optimal weights is performed, leading to
\begin{equation}
    S = -\frac{1}{2} \log(\det\mathbf{H}_o) + \mathrm{constants}.
\end{equation}
Here, as in Sec.~\ref{sec:uncertainty_estimate}, $\mathbf{H}_o$ is the Hessian matrix of the loss function at optimum. As previously discussed, the generalized Gauss-Newton approximation implies:
\begin{equation}
    \mathbf{H}_o \approx \mathbf{F}^\top \mathbf{F}
\end{equation}
Let us now take a new point of features $\mathbf{f}_\star$ and add it to the dataset. The new feature matrix $\mathbf{F}_\mathrm{new}$ is obtained by concatenating the row vector $\mathbf{f}_\star^\top$ to $\mathbf{F}$. The new Hessian becomes
\begin{equation}
    \mathbf{H}_{o, \mathrm{new}} \approx \mathbf{F}^\top \mathbf{F} + \mathbf{f}_\star \mathbf{f}_\star^\top
\end{equation}
and the new Shannon entropy is
\begin{equation}
\begin{split}
    S_{\mathrm{new}} &= -\frac{1}{2} \log(\det\mathbf{H}_{o,\mathrm{new}}) + \mathrm{constants} \\
    &\approx -\frac{1}{2} \log(\det(\mathbf{F}^\top \mathbf{F} + \mathbf{f}_\star \mathbf{f}_\star^\top)) + \mathrm{constants} \\
\end{split}
\end{equation}
One can use the \textit{matrix determinant lemma}\cite{vrabel2016note} to express:
\begin{equation}
    \det(\mathbf{F}^\top \mathbf{F} + \mathbf{f}_\star \mathbf{f}_\star^\top) = [1 + \mathbf{f}_\star^\top (\mathbf{F}^\top \mathbf{F})^{-1}\mathbf{f}_\star]\, \det(\mathbf{F}^\top \mathbf{F})
\end{equation}
The total information gain from adding $\star$ to the dataset, $\Delta I$, is
\begin{equation}
    \Delta I \equiv -(S_\mathrm{new} - S) = \frac{1}{2}\log[1 + \mathbf{f}_\star^\top (\mathbf{F}^\top \mathbf{F})^{-1}\mathbf{f}_\star]
\end{equation}
which is maximized when the (scaled) variance $\mathbf{f}_\star^\top (\mathbf{F}^\top \mathbf{F})^{-1}\mathbf{f}_\star$ is largest: therefore, to obtain maximal information gain (MIG), a next point should be chosen where the uncertainty, Eq.~\eqref{eq:sigma2star}, is currently largest. The MIG criterion has been recently used by K\"astner's group for active learning in atomistic simulations\cite{Zaverkin2024}. 
The MIG criterion also motivates, from an information theory perspective, the addition of structures characterized by environments with lowest \textit{local} prediction rigidity as active learning criterion, as in \textcite{chong2024rigidityfd}.
We remark that:
\begin{itemize}
    \item  the MIG criterion is \textit{independent of the specific target}, which need not be computed in advance to perform the active data selection.
    \item if we consider the initial dataset $\mathcal{D}$ as fixed, then maximizing the information gain implies looking for $\star$ to satisfy:
\begin{equation}
    \max_\star \det(\mathbf{F}_\mathrm{new}^\top \mathbf{F}_\mathrm{new})
\end{equation}
which is called D-optimality criterion in optimal design theory. $\mathbf{F}_\mathrm{new}^\top \mathbf{F}_\mathrm{new}$ is sometimes called, in this context, the \textit{Fisher information matrix} of the new dataset. We review the use of D-optimality for active learning in atomistic simulations in Sec.~\ref{sec:D-optimality}.
\item In this approach, the noise on data is taken the same for all the data (in fact, a single calibration constant $\alpha^2$ was used in Sec.~\ref{sec:uncertainty_estimate}). Generalization to the case of sample-dependent noise is nonetheless straightforward.
\end{itemize}

\subsubsection{D-optimality}\label{sec:D-optimality}
\textcite{Lysogorskiy2023} shows that the maximum deviation within an ensemble of models---in our notation $\max_i|\tilde{y}^{(i)} - \overline{y}|$, where $y$ is the total potential energy or a force component of a structure--- fully correlates with the so-called $D-$optimality criterion, which is then used for active learning strategy.
In the latter, one seeks to find 1) an optimal (sub)set of data samples and 2) to quantify how much a new sample $\star$ is represented. Specifically, if we collect all the dataset features in the ``tall'' $N_\mathrm{train}\times N_f$ matrix $\mathbf{F}$, step 1) looks at a subsampling $N_f$ data points, i.e.~selecting $N_f$ out of $N_\mathrm{train}$ rows to obtain a new, $N_f \times N_f$ matrix $\tilde{\mathbf{F}}$, such that
\begin{equation}
    \det({\tilde{\mathbf{F}}^\top \tilde{\mathbf{F}}})
\end{equation}
is maximal. 
Then, in step 2), a new sample $\star$ of features $\mathbf{f}_\star$ is selected from a pool of new data (e.g.~structures generated via MD trajectory) so that
\begin{equation}
    |\mathbf{f}_\star^\top \tilde{\mathbf{F}}^{-1}| = \sqrt{\mathbf{f}_\star^\top (\tilde{\mathbf{F}}^\top \tilde{\mathbf{F}})^{-1} \mathbf{f}_\star}
\end{equation}
is maximal (or larger than a given threshold $\gamma_\mathrm{th}\geq 1$). 

This is the same criterion found the previous section from the theory of maximal information gain, i.e. the quest for the sample $\star$ with largest variance, Eq.~\eqref{eq:sigma2star}. Nevertheless, this time, the adopted metric is $\tilde{\mathbf{G}} = (\tilde{\mathbf{F}}^\top \tilde{\mathbf{F}})^{-1}$, obtained with the D-optimally subsampled dataset.
Notice that
\begin{equation}
\begin{split}
    &\det(\tilde{\mathbf{F}}^\top \tilde{\mathbf{F}}) = \det(\tilde{\mathbf{F}})^2 \\
    &\leq \sum_{S} \det(\mathbf{F}_S)^2 = \det(\mathbf{F}^\top \mathbf{F}) = \lambda_1^2 \lambda_2^2 \cdots \lambda_{N_f}^2
\end{split}
\end{equation}
where $\lambda_i$ are the singular values of $\mathbf{F}$, $S$ runs over all the $\binom{N_\mathrm{train}}{N_f}$ combinations in which one can select $N_f$ rows out of $N_\mathrm{train}$ to build the $N_f \times N_f$ matrix $\mathbf{F}_S$. The second line follows from the Binet-Cauchy formula. A threshold on $\gamma$ must be set to determine whether a given  point is in- or out-of-distribution during the active learning cycle. The original paper by \textcite{Podryabinkin2017} suggests that a threshold of $\gamma \le1$ corresponds to prediction in in-distribution regime, and $\gamma \gg 1$ would correspond to strong out-of-distributions regimes.
D-optimal-based active learning for atomistic simulations and machine learning interatomic potentials construction has been implemented and extensively used first and foremost by the communities developing moment tensor potentials and atomic cluster expansion potentials \cite{Novikov2018, Podryabinkin2019,  Podryabinkin2022, Jalolov2024, Lysogorskiy2023}.
Just like the maximum-information-gain criterion, D-optimality proves to be largely more efficient than both random and CUR- (and FPS-) based selection \cite{Lysogorskiy2023}.

\subsubsection{Empirical forms of dataset entropy}\label{subsec:empirical_dataset_entropy}

Another approach that is used in the atomistic modeling community is based on the empirical estimate of the entropy of a distribution of dataset features. 
In this context, this set is usually taken as the set of features of \textit{atomic environments}\cite{karabin2020entropy,schwalbe2024information}. 
~\textcite{karabin2020entropy} propose the following estimator for the entropy of a distribution of features \textit{in a given configuration} $A$
(e.g., a given structure in a simulation cell) of $N_A$ atomic environments:
\begin{equation}
    S_\mathrm{KP}(A) = \frac{1}{N_A} \sum_{i=1}^{N_A} \log \left( N_A \min_j |\mathbf{x}_{A_i} - \mathbf{x}_{A_j}| \right)\label{eq:S_KP}
\end{equation}
where $|\mathbf{x}_{A_i} - \mathbf{x}_{A_j}|$ is the Euclidean distance between the atomic descriptors (features) of atoms $i$ and $j$.
This configuration-dependent entropy $S_\mathrm{KP}$ is then used for active dataset construction: the training set is incrementally built by adding independent local minima of the ``effective (free) energy'':
\begin{equation}
    V(A) = E_\mathrm{repulsive}(A) - K S_\mathrm{KP}(A)
\end{equation}
where $E_\mathrm{repulsive}$ is a short-range repulsive term penalizing very small distances between atoms, and $K$ is an entropy scaling coefficient which controls the relative importance of the two contributions. The minima are found via a simple annealing procedure at a given (fixed) cell volume.
Notice that one could also construct a \textit{global}  $S_\mathrm{KP}$ by considering \textit{all} the atomic environment features present in the dataset.
\footnote{\textcite{karabin2020entropy} report that \textit{``Extensions to global entropy-maximization over the whole training set (in contrast to the local configuration-by-configuration optimization presented here) are in development and will be reported in an upcoming publication.}'' ~\textcite{montes2022training} still adopts the local approach.}

Nonetheless, $S_\mathrm{KP}$ diverges to $-\infty$ whenever the features associated to environments $i$ and $j$ in Eq.~\eqref{eq:S_KP} coincide.
~\textcite{schwalbe2024information} solve this issue by defining a dataset entropy of a set of $N_\mathrm{env}$ environments:
\begin{equation}
    S_\mathrm{SK} = -\frac{1}{N_\mathrm{env}}\sum_{i=1}^{N_\mathrm{env}} \log\left[\frac{1}{N_\mathrm{env}} \sum_{j=1}^{N_\mathrm{env}} k(\mathbf{x}_i, \mathbf{x}_j) \right]
\end{equation}
where $k(\mathbf{x}_i, \mathbf{x}_j)$ is some kernel function expressing the similarity between environments $i$ and $j$. This formulation is also used to define a ``differential entropy'':
\begin{equation}
    \delta S_\mathrm{SK}(\mathbf{x}_\star) = -\log \left[\sum_{i=1}^{N_\mathrm{env}} k(\mathbf{x}_i,\mathbf{x}_\star) \right]\label{eq:S_SK_differential}
\end{equation}
which is then used for active learning and as a ``model-free uncertainty estimator'' of a new input for a given dataset\cite{schwalbe2024information}. 

We highlight that even when the dataset is built with targets that are structural properties, rather than local atomic properties, no ``grouping'' of local environments into structures is taken into account in these data-entropy based schemes, as it is instead done in the construction of the metric tensor in Mahalanobis distance (see Eq.~\eqref{eq:prloss}).
It is further relevant to note that any kernel can be written (Mercer's theorem) as
\begin{equation}
    k(\mathbf{x}_i, \mathbf{x}_j) = \sum_a \lambda_a \varphi_a(\mathbf{x}_i)\varphi_a(\mathbf{x}_j) =  \bm{\phi}^\top(\mathbf{x}_i)\, \bm{\phi}(\mathbf{x}_j)
\end{equation}
where $\lambda_a\geq 0$ and $\varphi_a$ are the eigenvalues and eigenfunctions of the kernel with respect to a measure $\mu$:
\begin{equation}
    \int k(\mathbf{x},\mathbf{x}') \varphi_a(\mathbf{x}') \mathrm{d}\mu(\mathbf{x}') = \lambda_a \varphi_a(\mathbf{x})
\end{equation}
and the (possibly infinite) components of $\bm{\phi}(\mathbf{x})$ are $\phi_a(\mathbf{x}) = \sqrt{\lambda_a} \varphi_a(\mathbf{x})$, for every possible $\mathbf{x}$. 
In such a case, 
\begin{equation}
    \sum_{i=1}^{N_\mathrm{env}}k(\mathbf{x}_i, \mathbf{x}_\star) = N_\mathrm{env} \overline{\bm{\phi}}^\top \, \bm{\phi}(\mathbf{x}_\star) 
\end{equation}
where $\overline{\bm{\phi}} = \frac{1}{N_\mathrm{env}} \sum_{i=1}^{N_\mathrm{env}} \bm{\phi}(\mathbf{x}_i)$ is the array of mean (latent) features over the dataset, i.e. the coordinates of the center of the dataset latent features. If we replace this into Eq.~\eqref{eq:S_SK_differential} we obtain, since the logarithm is a monotonic function, that the maximum differential entropy is given when the argument of the logarithm is minimal, i.e.
\begin{equation}
    \max_{\star}\left[\delta S_\mathrm{SK}(\mathbf{x}_\star) \right]\Leftrightarrow \min_{\star} \left[ \overline{\bm{\phi}}^\top \, \bm{\phi}(\mathbf{x}_\star) \right]
\end{equation}
Unfortunately, in contrast to the forms of active learning of the previous subsections, based on Eq.~\eqref{eq:sigma2star}, here the complexity of the dataset is effectively ``averaged out'' by considering $\overline{\bm{\phi}}$ in the latent feature space. Notice that if the features in the latent feature space are centered (see Appendix \ref{appendix:centering} for a discussion on whether latent feature centering is legitimate or not), $\overline{\bm{\phi}}$ vanishes, and one incurs into the additional problem that the argument of $\min_\star$ vanishes for any $\mathbf{x}_\star$.


A different perspective on assessing the proximity of two distribution (e.g., training points features and test point features) was proposed by~\textcite{Zeni2018}.
Their study considered 2-body machine learning potentials and the Kullback-Leibler divergence between distributions of interatomic distances in the training and test sets to rationalize prediction errors in machine learning potentials. 
The Kullback-Leibler divergence is an asymmetry statistical measure to quantify the information loss when a probability distribution $q(\xi)$ associated to a dataset $\mathcal{Q}$ over some sample space $\Xi$ is used to approximate another probability distribution $p(\xi)$ associated to a dataset $\mathcal{P}$ on the same sample space:
\begin{equation}
    D_{\text{KL}}(\mathcal{P}\| \mathcal{Q}) = \sum_{\xi\in\Xi} p(\xi) \log \frac{p(\xi)}{q(\xi)}
\end{equation}
A positive correlation between KL divergence and the mean absolute error of 2-body kernels was observed, highlighting the importance of including training data that captures interatomic distances relevant to the test set.
The KL divergence was thus proposed as a measure to assess how well structural features in the training dataset align with those in the test dataset and interpret model errors in a case study concerning machine learning potentials for Ni nanoclusters~\cite{Zeni2018}. 
Extending the assessment to 3-body machine learning potentials and the KL divergence of bond-angle distribution functions also  resulted in the observation of positive correlation between the two quantities.
We notice an hysteretical behavior exists in this metric, whereby the net cross-entropy change in first adding a point to the dataset and then removing it is nonzero (see Appendix \ref{appendix:hysteresis}).

\begin{figure*}
    \includegraphics[width=\textwidth]{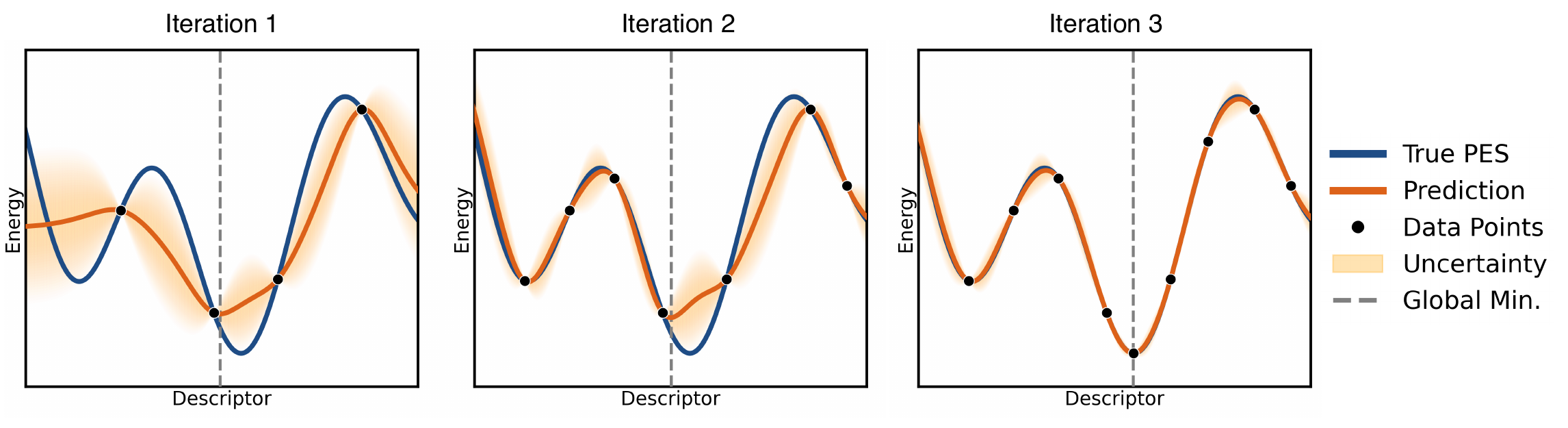}
    \caption{Example of Bayesian optimization in the sampling of a material energy landscapes.} 
    \label{fig:al}
\end{figure*}

\subsection{Bayesian Optimization} 
Bayesian optimization is a method designed to find the optimal value of a function efficiently. It uses a probabilistic model to predict the behavior of the function across the input space, guiding the search toward regions where the model is either uncertain or expects to find better results~\cite{mockus2013bayesian,shahriari2015taking,ramprasad2017machine,zhang2020bayesian}(Figure \ref{fig:al}).

While similar to active learning in its iterative approach and reliance on uncertainty to guide decisions, Bayesian optimization differs in its goal. Active learning focuses on generally improving a model's predictions. Bayesian optimization aims to optimize an objective function directly. 
Instead of stochastically sampling or evaluating all possibilities, Bayesian optimization identifies the next point to sample by balancing two goals: exploring unknown regions (where the function behavior is uncertain) and exploiting promising areas (where the function is predicted to perform well). Once the function is evaluated at the chosen point, the new information is used to update the model, and the process repeats until an optimal solution is found~\cite{snoek2012practical}.

Building upon these premises, Bayesian optimization acquisition deciding where to evaluate the objective function next and uncertainty estimates are at the heart of this process: e.g., 
Improvement\cite{jones1998efficient} uses uncertainty to identify areas where the potential for improvement is highest.
Probability of Improvement \cite{snoek2012practical} factors in uncertainty to assess the likelihood of finding better outcome.
The Upper Confidence Bound \cite{srinivas2009gaussian} takes a more explicit approach, blending the model's predictions with a weighted measure of uncertainty. 

Bayesian optimization has emerged as a powerful tool in atomistic modeling, offering efficient strategies for navigating complex energy landscapes and exploring vast configuration spaces. By leveraging probabilistic models, it enables the optimization of potential energy surfaces for intricate atomistic systems, guiding the search toward minima or other critical points with minimal computational cost.
Successful applications of BO in this area have been showcased for molecular~\cite{todorovic2019bayesian, shields2021bayesian}, crystalline~\cite{tran2020multi,yamashita2018crystal, zuo2021accelerating,ghorbani2024active} and disordered systems~\cite{shintaku2024first}, as well as complex interfaces \cite{ju2017designing}.

Beyond identifying stable configurations, Bayesian optimization facilitates the exploration of structures to achieve specific target properties or locate configurations along the Pareto front in multi-objective predictions. This capability makes it invaluable for applications ranging from material design to catalysis, where balancing multiple competing properties is often required.
Also in this case, numerous were the applications, including (but not limited to) metallic glasses mechanical properties \cite{makinen2024bayesian}, multi-principal~\cite{khatamsaz2023bayesian} or high-entropy alloys~\cite{torsti2024improving,kurunczi2024bayesian} and their catalytic properties~\cite{pedersen2021bayesian, ohyama2022bayesian, Janet2020}, or electrolytes properties for energy storage applications \cite{Jalem2018}.





\section{Uncertainty and data-efficient ML approaches}

Data-efficient methods offer a compelling pathway to achieve highly accurate prediction,  while significantly reducing the data and resource requirements. Notable emergent approaches in this area include Universal and Foundational Models, their fine-tuning, as well as Delta- and Multi-fidelity Learning.

Universal and foundational models (e.g., \cite{batatia2024, Deng2023}) are designed to capture broad physical relationships by training on diverse datasets. The accuracy of these models hinges on the large number of diverse training data; if critical subdomains are underrepresented, predictions in those regions may falter nevertheless.\cite{niblett2024transferabilitydatasetsmachinelearninginteraction}

Transfer learning builds upon the knowledge encoded in pre-trained models, such as universal and foundational models, adapting them to specific tasks or systems through fine-tuning with smaller datasets. This approach enhances data efficiency, and, often robustness, as the base model serves as a strong starting point. Nevertheless, challenges arise when the pre-trained model's domain significantly differs from the target domain.

Delta-learning \cite{Ramakrishnan2015} leverages a similar philosophy by predicting corrections on a simpler and (relatively) inexpensive model (e.g., a classical forcefield, a semi-empirical force-field, or a low quality DFT level). This method achieves high accuracy with minimal training data by concentrating on residual discrepancies. Also in this case, the quality of the baseline model and the representativeness of the training data are critical to Delta-learning model accuracy.

Multi-fidelity learning \cite{Allen2024} integrates information from datasets of varying accuracy and cost, effectively linking low-fidelity data to high-fidelity outputs. By synthesizing information from multiple sources, this approach enhances robustness while reducing the dependence on high-cost data. However, inconsistencies between fidelities and the challenge of accurately propagating uncertainties associated from models considering multiple fidelity levels demand careful consideration.


Open questions remain on how uncertainty estimate is affected by the use of these data-efficient models. These include (but are not limited to):
\begin{itemize}
    \item  \it{How do different baselines or fine-tuning strategies affect the reliability and robustness of the uncertainty estimators?}
    \item \it{Is the simultaneous learning of multiple level of theory advantageous in terms of both data efficiency and robustness ? How does it affect prediction uncertainty? }
\end{itemize}

\section{Conclusion}

In this perspective, we have examined the integration of machine learning and uncertainty quantification (UQ) in atomistic modeling, with a focus on methods to estimate uncertainties. 
We discussed state-of-the-art approaches, including Bayesian frameworks and ensemble techniques, and explored their applications in improving prediction reliability, guiding data acquisition through active learning and Bayesian optimization, and assessing the influence of uncertainties on equilibrium observables estimates. 
We also explored the influence of dataset composition and construction strategies on model accuracy, uncertainty, transferability, and robustness.
We finally considered emergent data-efficient approaches and highlighted emergent questions concerning prediction uncertainty estimate when leveraging these methods.

Taken together, our work underscores the role of rigorous UQ frameworks for guiding data-driven modeling and the value of thoughtful dataset construction in enhancing the transparency and robustness of ML-based atomistic modeling.
As new techniques—especially those geared toward data-efficient learning—continue to mature, careful validation and thorough uncertainty assessments become even more critical to maintain trust in model predictions.
We hope this perspective stimulates further development and integration of UQ protocols into atomistic modeling efforts.

Finally, we emphasize that the challenges and strategies for managing uncertainty in atomistic modeling echo a broader scientific discourse extending beyond this specific domain. 
Across materials science, physics, and chemistry, there is a renewed drive to establish clear standards for assessing information and uncertainty, from the reproducibility of synthetic protocols—whether in organic or materials synthesis—to the quality of data gleaned from real- and inverse-space characterization methods. 
The same principles underpin efforts to gauge the reliability of outputs from generative AI for materials discovery, large language models, automated image and spectrum analyses, and multi-modal approaches alike. 
Similar to Tycho Brahe's endeavor, these collective efforts will contribute to robust, transparent, and reproducible scientific findings.

%


\section{Acknowledgments}

We thank Alfredo Fiorentino, Matthias Kellner, Davide Tisi, and Claudio Zeni for fruitful discussions and critical comments to an early version of the manuscript. This article is based upon work from COST Action CA22154 - Data-driven Applications towards the Engineering of functional Materials: an Open Network (DAEMON) supported by COST
(European Cooperation in Science and Technology).
F.G. acknowledges financial support from UNIMORE through the FAR 2024 project “Revolutionizing All-Solid-State Sodium Batteries with Advanced Computational Tools and Mixed Glass Former Effects,” PI: Alfonso Pedone, CUP E93C24001990005, and from the EMPEROR Project, CUP E93C24001040001, sponsored by the National Quantum Science and Technology Institute (Spoke 5), Grant No. PE00000023, funded by the European Union -- NextGeneration EU.
S.C. acknowledges the support by the Swiss National Science Foundation (Project 200020\_214879).
S.B. thanks the European Union Horizon 2020 research and innovation program under grant agreement no.~857470 and from the European Regional Development Fund via the Foundation for Polish Science International Research Agenda PLUS program grant No.~MAB PLUS/2018/8. S.B. acknowledges the Foundation for Polish Science for the FIRST TEAM FENG.02.02-IP.05-0177/23 project.



\appendix
\section{Nystr\"om approximation}\label{appendix:Nystrom}
In the Nystr\"om approximations one considers a regression problem, Eq.~\eqref{eq:yGPR}, with $N_f$ equal to the number of sparse points, $M$, and where
\begin{equation}
    [\bm{\phi}(\mathbf{x})]^\top = \mathbf{k}(\mathbf{x}, \mathbf{X}_s)^\top \mathbf{U}_s \bm{\Lambda}_s^{-1/2}.
\end{equation}
In this formula, $\mathbf{k}(\mathbf{x}, \mathbf{X}_s)$ is the vector of the kernels between the input point $\mathbf{x}$ and each of the points in the sparse set, that are collected in the matrix $\mathbf{X}_s \in \mathbb{R}^{M\times D}$, while $\mathbf{U}_s \in \mathbb{R}^{M\times M}$ is the matrix of the eigenvectors of the sparse set kernel matrix, $\mathbf{K}_s \equiv \mathbf{K}(\mathbf{X}_s, \mathbf{X}_s)$, that has as entries the kernel between two points of the sparse set:
\begin{equation}
    \mathbf{K}_s = \mathbf{U}_s \bm{\Lambda}_s \mathbf{U}_s^\top.
\end{equation}
The diagonal matrix $\bm{\Lambda}$ collects the eigenvalues, ordered to correspond to $\mathbf{U}_s$.
The kernel matrix of the training set is then approximated as (Nystr\"om formula):
\begin{equation}
\begin{split}
    &\mathbf{K}(\mathbf{X}, \mathbf{X}) \approx \mathbf{K}(\mathbf{X}, \mathbf{X}_s) \left[\mathbf{K}_s\right]^{-1} \mathbf{K}(\mathbf{X}, \mathbf{X}_s)^\top \\
    &= \mathbf{K}(\mathbf{X}, \mathbf{X}_s) \mathbf{U}_s \bm{\Lambda}_s^{-1/2} \bm{\Lambda}_s^{-1/2} \mathbf{U}_s^\top \mathbf{K}(\mathbf{X}, \mathbf{X}_s)^\top \\
    &= \bm{\Phi} \bm{\Phi}^\top    
\end{split}
\end{equation}
Here, $\mathbf{X}\in \mathbb{R}^{N_\mathrm{train}\times D}$ is the training set matrix, and $\mathbf{K}(\mathbf{X}, \mathbf{X}_s)\in \mathbb{R}^{N_\mathrm{train}\times M}$ is the kernel matrix between the training set points and the sparse set points. After centering the $\mathbf{\Phi}$ matrix, the variance on the prediction for input $\star$ is readily obtained as Eq.~\eqref{eq:GPR_cov}.

\section{Hysteresis of cross-entropy gain/loss}\label{appendix:hysteresis}

Consider an initial dataset $ \mathcal{Q} $ and a probability density $q$ associated to it, then add a data point to obtain the distribution $ p $ associated with the new dataset $ \mathcal{P} $. The Kullback-Leibler (KL) divergence is given by:
\begin{equation}
D_\mathrm{KL}(\mathcal{P} \| \mathcal{Q}) = \sum_{\xi\in \Xi} p(\xi) \log\left(\frac{p(\xi)}{q(\xi)}\right)
\end{equation}
Then, starting from the dataset $ \mathcal{P} $, remove a data point to return to $ \mathcal{Q} $. The KL divergence in this case is:
\begin{equation}
D_\mathrm{KL}(\mathcal{Q} \| \mathcal{P}) = \sum_{\xi\in\Xi} q(\xi) \log\left(\frac{q(\xi)}{p(\xi)}\right)
\end{equation}
In general, $D_\mathrm{KL}(\mathcal{Q} \| \mathcal{P})\neq - D_\mathrm{KL}(\mathcal{P} \| \mathcal{Q})$. This results in a form of information hysteresis in the cycle $\mathcal{Q}\to \mathcal{P}\to \mathcal{Q}$, if we associate the KL divergence with the concept of information gain or loss. Notation was kept loose on purpose: Information hysteresis would exist irrespective of whether $p$ and $q$ represent (posterior) probability distribution of the weights,\cite{mackay1992information} as in Sec.~\ref{subsec:information_gain}, or the probability distribution of dataset features, as in Secs.~\ref{subsec:empirical_dataset_entropy}.

\section{Why (not) center (pseudo)features?}\label{appendix:centering}

In the discussion above, where the uncertainty of a prediction was interpreted as a Mahalanobis distance, we assumed that the distribution of the (pseudo)features was centered at zero. In linear regression, centering the features ensures that the intercept has a meaningful interpretation, such as representing the mean response when all predictors are at their mean values. 
In kernel methods, however, the focus shifts to pairwise similarities encoded in the kernel matrix, which implicitly maps the data into a latent space of pseudofeatures. 

Centering the kernel—either directly or by centering the pseudofeatures as in the Nyström approximation—adjusts the distribution of data representation in latent space, thereby affecting the variance of predictions, which may reflect both the global mean effect and deviations from this mean. 
Centering also isolates variability purely due to deviations, aligning the variance estimate more closely with the concept used in linear models, where centering is standard. 

\begin{figure}
    \centering
    \includegraphics[width=0.85\columnwidth]{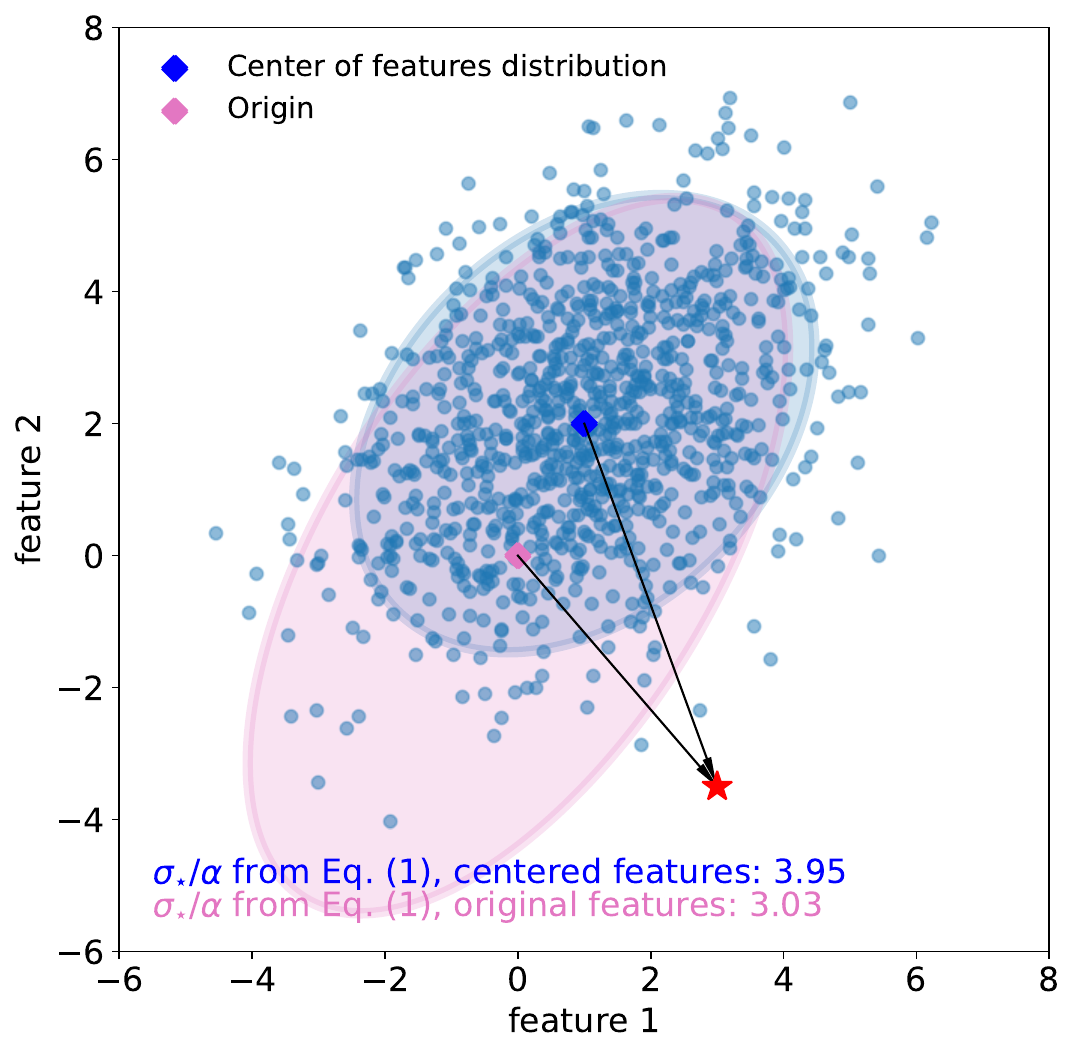}
    \caption{Effect of application of Eq.~\eqref{eq:sigma2star}, where $\mathbf{f}_\star$ and $\mathbf{F}$ are centered or not with respect to the center of the dataset's features distribution, for a toy model with two features. The (unitless) Mahalanobis distance is obtained by expressing the uncertainty from Eq.~\eqref{eq:sigma2star} in units of calibration parameter $\alpha$.}
    \label{fig:Mahalanobis_shifted}
\end{figure}
However, pseudofeatures centering has a direct impact on variance estimates, especially for bias-less models. For instance, consider a NN model where the output $y_\star = \mathbf{f}_\star^\top \mathbf{w}^L$ is forced to vanish for vanishing features no matter the values assumed by the last-layer weights: the uncertainty on the prediction, $\sigma_\star(\mathbf{f}_\star = 0)$, is always zero by construction---see Eq.~\eqref{eq:sigma2star}. In general, the variance for predictions near (far from) the origin of the latent space will be small (large). Nonetheless, this is a characteristic of the model: one could in fact argue that re-centering may introduce spurious \textit{a posteriori} effects that clash with how the model ultimately represents (or learns) data in latent space. 
By centering the input features one effectively removes the global mean effect, ensuring that the predictions reflect deviations based solely on the relative relationships between data points.
Yet, it is not evident that performing centering on pseudofeatures (that are the way the kernel represents data, or are learned by the model in NN architectures) should be encouraged.

\bibliography{biblio_full}


\end{document}

%% file: preamble.tex
\usepackage{graphicx}
\usepackage[colorlinks=true, linkcolor=green, citecolor=green]{hyperref}
\usepackage{amsfonts}
\usepackage{amsmath}
\usepackage{bm}

\usepackage{xcolor}
\usepackage[normalem]{ulem}
\definecolor{tangerine}{rgb}{0.944,0.522,0}
\definecolor{verde}{rgb}{0.,0.6,0}
\definecolor{rosso}{rgb}{0.9,0.0,0.2}
\definecolor{magenta}{rgb}{0.9,0.2,0.9}

\newif\ifhighlight

\newcommand{\highlight}{\highlighttrue}
\highlight
\newcommand{\editor}[2]{%
  \expandafter\newcommand\csname #1note\endcsname[1]{%
    \textcolor{#2}{(\textbf{#1note:} \textsc{##1})}}%
  \expandafter\newcommand\csname #1\endcsname[1]{%
    \ifhighlight\textcolor{#2}{##1} \else ##1\fi}%
  \expandafter\newcommand\csname #1cancel\endcsname[1]{%
    \ifhighlight\textcolor{#2}{\sout{##1}}\fi}%
  \expandafter\newcommand\csname #1change\endcsname[2]{%
    \ifhighlight\textcolor{#2}{\sout{##1} ##2}\else ##2\fi}%
  \newenvironment{#1text}{\ifhighlight\color{#2}\fi}{\color{black}}
}

\editor{FG}{rosso}
\editor{KR}{tangerine}
\editor{SC}{purple}